\documentclass[twocolumn,trackchanges]{aastex701}
\usepackage{amsmath}
\usepackage{natbib}

\begin{document}

\title{First Statistical Detection of \ion{Mg}{2}-traced Cool Gas Outflows with JWST Towards Cosmic Dawn}

\author[orcid=0009-0000-7307-6362]{Cheqiu Lyu}
\altaffiliation{These authors contributed equally to this work.}
\affiliation{Department of Astronomy, University of Science and Technology of China, Hefei, Anhui 230026, China}
\affiliation{School of Astronomy and Space Science, University of Science and Technology of China, Hefei, Anhui 230026, China}
\email[show]{lyucq@ustc.edu.cn}  

\author[orcid=0009-0008-1319-498X]{Haoran Yu} 
\altaffiliation{These authors contributed equally to this work.}
\affiliation{Department of Astronomy, University of Science and Technology of China, Hefei, Anhui 230026, China}
\affiliation{School of Astronomy and Space Science, University of Science and Technology of China, Hefei, Anhui 230026, China}
\email{ryanr@mail.ustc.edu.cn}

\author[orcid=0000-0003-1588-9394]{Enci Wang}
\affiliation{Department of Astronomy, University of Science and Technology of China, Hefei, Anhui 230026, China}
\affiliation{School of Astronomy and Space Science, University of Science and Technology of China, Hefei, Anhui 230026, China}
\email[show]{ecwang16@ustc.edu.cn}

\author[0000-0002-4419-6434]{Junxian Wang}
\affiliation{Department of Astronomy, University of Science and Technology of China, Hefei, Anhui 230026, China}
\affiliation{School of Astronomy and Space Science, University of Science and Technology of China, Hefei, Anhui 230026, China}
\email[]{XXX}

\author[0009-0004-7042-4172]{Cheng Jia}
\affiliation{Department of Astronomy, University of Science and Technology of China, Hefei, Anhui 230026, China}
\affiliation{School of Astronomy and Space Science, University of Science and Technology of China, Hefei, Anhui 230026, China}
\email[]{XXX}

\author[0000-0002-0846-7591]{Jie Song}
\affiliation{Department of Astronomy, University of Science and Technology of China, Hefei, Anhui 230026, China}
\affiliation{School of Astronomy and Space Science, University of Science and Technology of China, Hefei, Anhui 230026, China}
\email[]{XXX}

\author[0000-0002-4597-5798]{Yangyao Chen}
\affiliation{School of Astronomy and Space Science, Nanjing University, Nanjing, Jiangsu 210093, China}
\affiliation{Key Laboratory of Modern Astronomy and Astrophysics, Nanjing University, Ministry of Education, Nanjing, Jiangsu 210093, China}
\email[]{XXX}

\author{Jinyang Wang}
\affiliation{Department of Astronomy, University of Science and Technology of China, Hefei, Anhui 230026, China}
\affiliation{School of Astronomy and Space Science, University of Science and Technology of China, Hefei, Anhui 230026, China}
\email[]{XXX}

\author[0009-0004-5989-6005]{Zeyu Chen}
\affiliation{Department of Astronomy, University of Science and Technology of China, Hefei, Anhui 230026, China}
\affiliation{School of Astronomy and Space Science, University of Science and Technology of China, Hefei, Anhui 230026, China}
\email[]{XXX}

\author[0009-0006-7343-8013]{Chengyu Ma}
\affiliation{Department of Astronomy, University of Science and Technology of China, Hefei, Anhui 230026, China}
\affiliation{School of Astronomy and Space Science, University of Science and Technology of China, Hefei, Anhui 230026, China}
\email[]{XXX}

\author{Yifan Wang}
\affiliation{Department of Astronomy, University of Science and Technology of China, Hefei, Anhui 230026, China}
\affiliation{School of Astronomy and Space Science, University of Science and Technology of China, Hefei, Anhui 230026, China}
\email[]{XXX}

\author[0000-0002-7660-2273]{Xu Kong}
\affiliation{Department of Astronomy, University of Science and Technology of China, Hefei, Anhui 230026, China}
\affiliation{School of Astronomy and Space Science, University of Science and Technology of China, Hefei, Anhui 230026, China}
\email[]{XXX}
\begin{abstract}

Galactic-scale outflows are a crucial component of galaxy evolution, yet their properties in the early universe remain poorly constrained. We present the first statistical investigation of \ion{Mg}{2}-traced cool gas outflows in galaxies spanning a wide cosmic timeline from $z \approx 1$ to $z > 6$ (with sample coverage extending to $z \sim 10$). Using thousands of public JWST/NIRSpec spectra, we employ a signal-to-noise weighted spectral stacking technique on the \ion{Mg}{2} $\lambda\lambda2796, 2803$ absorption doublet. We robustly detect blueshifted \ion{Mg}{2} absorption in nearly all stellar mass and redshift bins, with the exception of the lowest-mass systems at $z \sim 1-2$. The outflow equivalent width exhibits a positive correlation with stellar mass ($M_*$) at all epochs, with the fitted slope of $1.21 \pm 0.35$. Our work provides the first statistical constraints on \ion{Mg}{2}-traced cool outflows in the low-mass ($M_* \lesssim 10^{9.5} \, \mathrm{M}_\odot$), high-redshift ($z > 3$) regime. We also find that the outflow velocities generally remain below the host halo escape velocities, consistent with a galactic fountain scenario. The consistency of the stellar mass-outflow equivalent width relation across $z \sim 2-6$ suggests a persistent, unevolving feedback mechanism governing the baryon cycle towards cosmic dawn.

\end{abstract}


\section{Introduction} 

Galactic-scale outflows are fundamental to our understanding of galaxy evolution. Driven by energetic feedback from massive stars, supernovae (SNe), and active galactic nuclei (AGN) \citep[e.g.,][]{Silk1998, Veilleux2005, Croton2006, Veilleux2020}, these outflows act as a key regulatory agent in the baryon cycle between galaxies and their surrounding environments. By expelling gas from the interstellar medium (ISM), they modulate the fuel available for star formation, thereby shaping the mass assembly of galaxies over cosmic time \citep[e.g.,][]{DiMatteo2005, Hopkins2008, Somerville2008, Erb2015, Wang-2019}. Furthermore, these outflows are the primary mechanism for transporting chemically enriched material from galaxies into the circumgalactic medium (CGM) and even the intergalactic medium (IGM), playing a crucial role in the cosmic metal enrichment history \citep[e.g.,][]{Pettini2001, Tremonti2004, Finlator2008, Tumlinson2017, Wang2021, Wang-2022a, Ma-2024, Chen2025a, Chen2025b, Jia-2025, Lyu2025a}. Despite their acknowledged importance in theoretical models and hydrodynamical simulations of galaxy formation, establishing robust observational constraints on the physical properties of outflows remains a significant challenge, particularly in the early universe.

Observing outflows becomes progressively more difficult at high redshifts. During the peak epoch of cosmic star formation ($z \sim 1-3$), outflows are known to be ubiquitous in massive star-forming galaxies \citep[e.g.,][]{Weiner2009, Rubin2014, ForsterSchreiber2019}. However, pushing these studies towards the epoch of cosmic dawn ($z > 6$) presents formidable challenges. The intrinsic faintness of high-redshift galaxies makes it extremely difficult to obtain spectra with sufficient signal-to-noise ratio (S/N) to detect the subtle absorption features imprinted by outflows on the galaxy's continuum. 

Despite these obstacles, significant progress has been made in characterizing outflows beyond $z \sim 3$ using various facilities. \citet{Sugahara2019} utilized deep ground-based spectra to identify outflows via low-ionization rest-frame UV absorption lines in composite spectra of galaxies at $z=5-6$. In the sub-millimeter regime, \citet{Birkin2025} employed ALMA stacking of [C~II] 158$\mu$m emission to probe cold gas outflows at $z \sim 5$. More recently, the advent of the James Webb Space Telescope \citep[JWST,][]{Gardner2006} has revolutionized this field. With its unprecedented sensitivity and near-infrared spectroscopic capabilities, JWST/NIRSpec enables the detailed probing of rest-frame optical features that were previously inaccessible. Through the analysis of broad components in emission lines (e.g., H$\alpha$, [O~III]), several studies have statistically characterized ionized outflows in galaxies spanning $z \sim 3$ to $z \sim 9$ \citep{Carniani2024, Saldana-Lopez2025, Cooper2025, Xu2025}. These works generally confirm that feedback mechanisms are active in the early universe, yet they primarily focus on the ionized gas phase.

Absorption line spectroscopy is also a powerful and widely used technique to probe the kinematics of cool gas flows along the line of sight. Blueshifted low-ionization metal absorption lines, such as the \ion{Mg}{2} $\lambda\lambda2796, 2803$ doublet and the \ion{Na}{1} D $\lambda\lambda5890, 5896$ doublet, are particularly effective tracers of cool, neutral gas in outflows \citep[e.g.,][]{Weiner2009, Martin2012, Rubin2014}. The \ion{Mg}{2} doublet, commonly detected in the CGM of galaxies, is a robust indicator of enriched cool gas ($T \sim 10^4$ K) over a wide range of \ion{H}{1} column densities \citep{Bergeron1986, Churchill2000}. The \ion{Na}{1} D doublet, with its low ionization potential (5.1 eV), specifically traces dense, neutral gas ($T < 10^3$ K) often shielded by dust \citep{Savage1996, Puspitarini2012, Baron2020}. However, detecting these subtle absorption features in individual high-redshift galaxies remains challenging. \citet{Valentino2025} recently reported individual detections of blueshifted \ion{Mg}{2} absorption in two rare massive quiescent galaxies at $z \sim 4$ and $z \sim 7$. However, a statistical census of the cool, low-ionization gas phase traced by \ion{Mg}{2} in the typical, abundant star-forming galaxy population remains largely unexplored.

JWST/NIRSpec enables the detection of rest-frame optical features like \ion{Na}{1} D at high redshifts for the first time \citep[e.g.,][]{Belli2024, Cresci2023, Davies2024, Kehoe2025}, and UV tracers like \ion{Mg}{2} across a vast cosmic timeline ($z \sim 1-10$). The ability to assemble large spectroscopic samples of typical galaxies at these early epochs finally opens the door to characterizing the statistical properties of outflows and establishing crucial scaling relations that link outflow properties to intrinsic galaxy parameters, such as stellar mass and redshift.

JWST also opens a new frontier for testing galaxy formation theories in the early universe. Recently, new models have been proposed to explain the surprising abundance of luminous galaxies observed by JWST at $z \gtrsim 10$. One prominent example is the ``feedback-free starburst" (FFB) model \citep{Dekel2023, Li2024}. This model posits that at the high gas densities and low metallicities prevalent at cosmic dawn, the star formation timescale can become shorter than the delay time for stellar feedback from supernovae and stellar winds. This could lead to a highly efficient, bursty mode of star formation, fundamentally different from the self-regulating feedback processes observed at later cosmic times. The physical properties of galactic outflows—such as their velocities, equivalent widths, and scaling relations with galaxy properties—serve as direct probes of the underlying feedback mechanisms. By statistically characterizing outflows towards cosmic dawn, we can test the predictions of models like the FFB and determine whether a shift in the nature of feedback indeed occurred in the early universe.

In this work, we leverage the vast public dataset from the DAWN JWST Archive (DJA) to conduct the statistical investigation of cool gas outflows in galaxies spanning the redshift range $1 < z < 10$. Note that while our sample includes galaxies out to $z \approx 10$, the S/N-weighted stacking in the highest redshift bin is most sensitive to the population at $z \sim 6$, see Table \ref{tab:table} for weighted mean redshifts. By employing a spectral stacking technique on hundreds of JWST/NIRSpec spectra, we significantly enhance the S/N, enabling the detection of the faint, average outflow signals that are invisible in individual spectra. The primary goals of this Letter are: (1) to reliably detect outflow traced by \ion{Mg}{2} and \ion{Na}{1} D at different redshift, and measure the average outflow velocities and equivalent widths traced by \ion{Mg}{2} across a range of redshifts and stellar masses; and (2) to quantitatively establish the scaling relations between outflow properties, galaxy stellar mass, and across cosmic time; and (3) to provide key observational constraints on the physics of galactic feedback and its role in galaxy evolution towards the cosmic dawn.

This Letter is structured as follows. In Section~\ref{sec:data}, we describe our data and sample selection. In Section~\ref{sec:method}, we detail our methodology for spectral stacking and measuring outflow properties. We present our main results in Section~\ref{sec:results} and discussion in Section~\ref{sec:discussion}. Finally, we summarize our findings in Section~\ref{sec:summary}. Throughout this Letter, we adopt a $\Lambda$CDM cosmology with $H_0 = 70$ km s$^{-1}$ Mpc$^{-1}$, $\Omega_m = 0.3$, and $\Omega_\Lambda = 0.7$, and assume a \citet{Chabrier2003} stellar initial mass function (IMF). We refer to the doublets by their conventional names (e.g., \ion{Mg}{2} $\lambda\lambda$2796, 2803 and \ion{Na}{1} D $\lambda\lambda$5890, 5896). However, all calculations are performed using the precise vacuum rest-frame wavelengths from \citet{Morton2003}: $\lambda_{\rm vac} = 2796.3543, 2803.5315$\,\AA\ for \ion{Mg}{2}, and $\lambda_{\rm vac} = 5891.5833, 5897.5581$\,\AA\ for \ion{Na}{1} D, consistent with the JWST spectral calibration.

\section{Data and Sample Selection} \label{sec:data}

\begin{figure*}[ht!]
\includegraphics[width=0.33\linewidth]{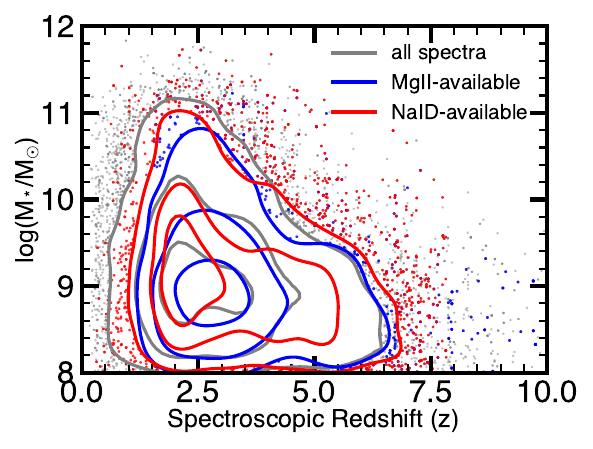}
\includegraphics[width=0.33\linewidth]{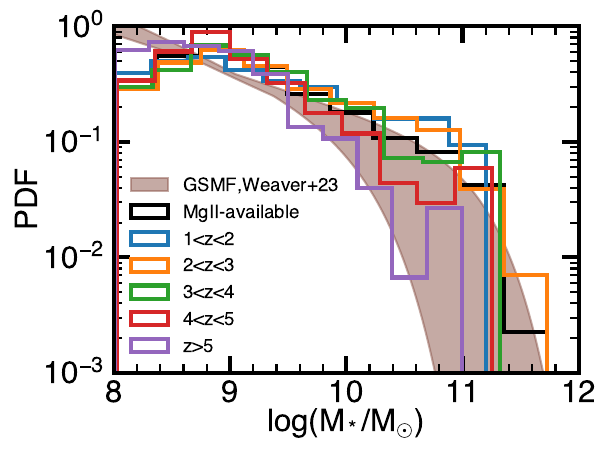}
\includegraphics[width=0.33\linewidth]{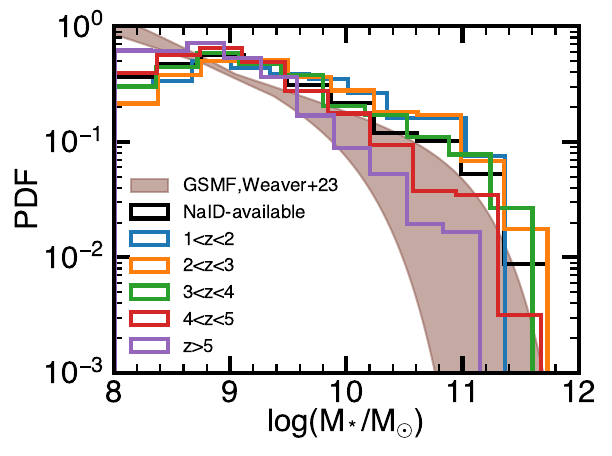}
\caption{Sample selection and stellar mass distribution. The left panel displays the distribution of our galaxy samples in the stellar mass-redshift plane. The grey contours and scatter points represent the parent sample of 24,927 galaxies, while the blue and red contours and points show the subsamples with valid spectral coverage of the \ion{Mg}{2} and \ion{Na}{1}~D doublets, respectively. The three contour levels for each sample are derived from a Kernel Density Estimate (KDE) and correspond to iso-density boundaries enclosing 16\%, 50\%, and 84\% of the distribution. Individual galaxies located in low-density regions (outside the 84\% contour) are plotted as scatter points to illustrate the full extent of the sample. The middle and right panels present the Probability Density Functions (PDFs) of the stellar mass distributions for the \ion{Mg}{2}- and \ion{Na}{1}~D-available subsamples. The colored histograms correspond to the different redshift bins as labeled. To assess the representativeness of our sample, we overlay the evolving Galaxy Stellar Mass Function (GSMF) from \citet{Weaver2023}. The brown shaded region represents the envelope of the GSMF across the redshift range $1.1 < z < 7.5$, derived from the Schechter function parameters (double for $z<3$ and single for $z>3$) presented in their work.
\label{fig:sample}}
\end{figure*}

Our analysis is based on the JWST/NIRSpec spectra dataset from the DAWN JWST Archive (DJA)\footnote{\url{https://dawn-cph.github.io/dja/}}. Specifically, we utilize the v4.4 Merged Table \footnote{\url{https://dawn-cph.github.io/dja/blog/2025/05/01/nirspec-merged-table-v4/}} and its associated extended spectra catalog \citep{Valentino2025, Pollock2025}. This comprehensive public catalog provides 80,367 uniformly processed spectra, reduced using the \texttt{msaexp} \citep{Brammer2022, deGraaff2025, Heintz2025} and \texttt{grizli} \citep{Brammer2023} pipelines. The data products include 1D/2D spectral extractions, redshift measurements, emission-line fluxes, and associated photometry. The complete dataset is available at DOI: \href{https://doi.org/10.5281/zenodo.1547235}{10.5281/zenodo.1547235}.

We construct our parent sample by selecting galaxies with a photometrically derived stellar mass ($M_{*}$) greater than $10^8~M_{\odot}$ and a secure spectroscopic redshift. The redshift quality is assessed via visual inspection flags provided in the catalog. We exclude spectra with poor quality flags, specifically \texttt{Grade}$=0$ (spectrum suffers some data quality issue) and \texttt{Grade}=$-1$ (fit not performed or graded). We note that the vast majority (approximately 99.5\%) of our selected galaxies are assigned \texttt{Grade 3}, indicating robust redshifts derived from multiple high-S/N emission lines. For these sources, the typical redshift uncertainty corresponds to a velocity error of $\lesssim 30$ km s$^{-1}$, which is negligible compared to the typical outflow velocities derived in this work. This initial selection results in a sample of 24,927 galaxy spectra.

To search for cool gas outflows, we create subsamples based on the spectral coverage of the \ion{Mg}{2} $\lambda\lambda2796, 2803$ and \ion{Na}{1} D $\lambda\lambda5890, 5896$ doublets. We select objects observed with NIRSpec medium- or high-resolution grisms (G140M/H, G235M/H, G395M/H; $R \sim 1000/2700$) that provide unmasked spectral coverage of a 100~\AA\ window roughly centered on these doublets (i.e., 2750--2850~\AA\ for \ion{Mg}{2} and 5850--5950~\AA\ for \ion{Na}{1} D). This selection ensures that the absorption features, if present, fall within the observed spectral range and are sampled at a sufficient resolution for detection. This procedure yields final samples of approximately 2,394 spectra for the \ion{Mg}{2} analysis and 5,512 spectra for the \ion{Na}{1} D analysis. However, due to the emission line blending and inherent stellar contamination, we do not investigate the scaling relations with galaxy properties of \ion{Na}{1}~D sample in this work (See Section \ref{sec31} for details).

The properties and representativeness of our final samples are summarized in Figure~\ref{fig:sample}. The left panel shows the coverage of our parent sample and the \ion{Mg}{2}- and \ion{Na}{1}~D-available subsamples across the stellar mass-redshift plane. The middle and right panels show the normalized stellar mass distributions (as PDFs) for our subsamples, binned by redshift. 

To investigate whether our sample is a representative galaxy sample, we compare our distributions to the evolving Galaxy Stellar Mass Function (GSMF) from the comprehensive JWST analysis of \citet{Weaver2023}. Although our sample is from a wide range of programs and surveys for different proposes, as shown by the brown shaded envelope in Figure~\ref{fig:sample}, which encapsulates the GSMF evolution from $z=1.1$ to $z=7.5$, our sample distributions trace the shape of the underlying galaxy population remarkably well, particularly at the high-mass end ($M_* > 10^9 \rm{M_\odot}$) where the majority of the cosmic stellar mass resides. This agreement confirms that our spectroscopic sample is broadly representative of the typical galaxy population across the full mass and redshift range of our study. Therefore, the outflow properties derived in this work are characteristic of the general galaxy population during these cosmic epochs.

\section{Spectra Stacking Method}\label{sec:method}

To systematically investigate the prevalence of galactic outflows across the galaxy population, we stack galaxy spectra binned by their physical properties.
This technique significantly improves S/N, enabling the detection of outflow signatures that are too faint to identify in individual spectra \citep[e.g.][]{Weiner2009, Chen2010, Rubin2010, Bordoloi2014}.

\subsection{Weighted Median Stacking}

We begin by shifting all individual spectra to their rest-frame.
For each spectrum, we extract the spectral data within a 200\AA\ window centered on the doublet features (2700--2900\AA\ for \ion{Mg}{2} and 5800--6000\AA\ for \ion{Na}{1} D).
These segments are then resampled onto a unified wavelength grid using \texttt{Spectres} \citep{Carnall2017}.
To avoid artificial super-sampling, which could introduce additional errors, the wavelength interval for the resampled spectra is inferred from the sample redshift, resulting in adopted intervals of $\Delta \lambda =1.4\ {\rm\AA}, 0.9\ {\rm\AA}, 0.7\ {\rm\AA}, 0.6\ {\rm\AA}, 0.5\ {\rm\AA}$ for $1<z<2$, $2<z<3$, $3<z<4$, $4<z<5$ and $5<z<10$ bins, respectively.
After masking the prominent lines (\ion{Mg}{2}, \ion{Mg}{1}, \ion{Na}{1} D), we estimate the continuum by first applying a median filter with an 11-pixel window and then fitting a 5th-order polynomial to the smoothed spectrum. 
Each original spectrum is divided by its corresponding continuum to produce a normalized spectrum.
In idealized circumstances, the normalized flux should be unity across non-absorbing regions.
However, the normalization can be biased in cases of low S/N or unexpected strong absorption features, thereby introducing additional uncertainty into the stacked spectrum.
For instance, if the fitted continuum approaches or crosses zero, the normalized flux can diverge, producing a singularity.

Aiming at minimizing the impact of poorly constrained normalizations to the spectra stacking processes, we define a weight, $w$, to evaluate the reliability of the spectra normalization.
It is defined as
\begin{equation}\label{eq:weight1}
    w\equiv \frac{1}{\sum_{i=1}^N (F_i-F_i^{\rm ref})^2 / N},
\end{equation}
where $F_i$ is the $i$th flux density in the normalized spectrum, $F_i^{\rm ref}=1$ is the reference value for a perfectly normalized spectrum, and $N$ is the number of the spectral data points.
In this calculation, the regions with expected spectral lines (\ion{Mg}{2}, \ion{Mg}{1}, \ion{Na}{1} D) are masked. We then generate the median-stacked spectra based on these weights.
The composite flux in each wavelength bin is computed as the weighted median of all individual fluxes.
For $n$ ordered fluxes $f_1,f_2,\cdots,f_n$ with corresponding weights $w_1,w_2, \cdots,w_n$, the weighted median is defined as the element $f_k$ satisfying 
\begin{equation}\label{eq:weight2}
    \frac{\sum_{i=1}^{k-1}w_i}{\sum_{i=1}^{n}w_i} \leq \frac{1}{2}\quad {\rm and}\quad 
    \frac{\sum_{i=k+1}^{n}w_i}{\sum_{i=1}^{n}w_i} \leq \frac{1}{2}.
\end{equation}
In such a regime, the impact of the spectra with poor normalization is minimized, because their weights are extremely small.
Spectra of this kind occupy approximately 50\% of the sample.
Furthermore, to prevent a small fraction of spectra with high weights dominating the stack, we impose an upper threshold $w^{\rm thr}$, thereby
\begin{equation}\label{eq:weight3}
    w = \left\{\begin{aligned}
        &w, &\quad {\rm if}\ w< w^{\rm thr},\\
        &w^{\rm thr},&\quad {\rm otherwise}.
    \end{aligned}
    \right.
\end{equation}
Based on inspection of the weight distribution, we adopt $w^{\rm thr}=1$ to mitigate the potential bias from the high-weight tail, and this threshold is employed throughout our study.

To quantify the uncertainty in the final stacked spectra, we employ a bootstrap resampling procedure.
Specifically, we generate 1000 bootstrap samples, each by randomly drawing $N$ spectra from the original set of $N$ with replacement.

For comparison, we have also performed an unweighted stacking analysis with equal probabilities; the results are presented in the Appendix \ref{sec:app}.

When determining the representative $z$, $M_*$ of a bin, we employ the same weighting method, e.g., $\langle z\rangle_{w}$ is derived as the weighted median redshift using exactly the same weights derived from Equation \ref{eq:weight1}, \ref{eq:weight2} and \ref{eq:weight3}.

\subsection{Outflow Properties}

We employ the \texttt{boxcar} method \citep{Rubin2010, Bordoloi2014} to derive the mean outflow velocity ($v_{\rm out}$) and the outflow equivalent width ($\rm EW_{out}$) of the \ion{Mg}{2} doublet.
This approach provides model-independent measurements directly from the observed absorption profile, which is advantageous for our statistical analysis.
Given the blending between the red-wing of \ion{Mg}{2}\ $\lambda$2796 line and the blueshifted \ion{Mg}{2}\ $\lambda$2803 absorption, we use the blue-wing of \ion{Mg}{2}\ $\lambda$2796 and the red-wing of \ion{Mg}{2}\ $\lambda$2803 to trace the kinematics of the species.
We conceptually decompose the absorption into a blueshifted outflowing component (denoted as ``out'') characterized by a blueshifted line center, and an interstellar component (denoted as ``sym'') with a symmetrical profile centered at the rest-frame wavelength.
The EWs are measured in two velocity intervals: $\rm EW_{blue}$ from $-700$ to $0~\rm km\ s^{-1}$ relative to \ion{Mg}{2}\ $\lambda$2796, and $\rm EW_{red}$ from $0$ to $+400~\rm km\ s^{-1}$ relative to \ion{Mg}{2}\ $\lambda$2803, which are defined empirically to cover the extended absorption wings.
Following \citet{Bordoloi2014}, the total absorption EW, defined as $\rm EW_{tot}\equiv EW_{out}+EW_{sym}$, and the outflow EW are estimated as
\begin{equation}
    \begin{aligned}
        \rm EW_{tot}&=\rm 2(EW_{blue}+EW_{red}),\\
        \rm EW_{out}&=\rm2(EW_{blue}-EW_{red}).
    \end{aligned}
\end{equation}
In this analysis, the absorption profiles of the doublet are assumed to be identical. 
This assumption is supported by observational evidence that the doublet ratio in most \ion{Mg}{2} absorbers is close to 1:1 due to saturation, indicating that the cool, neutral phase gas traced by \ion{Mg}{2} is clumpy and optically thick \citep[e.g.,][]{Lan17}.
Accordingly, the mean velocity of a single absorption trough is derived by merging the spectrum blueward of \ion{Mg}{2}~$\lambda$2796 with that redward of \ion{Mg}{2}~$\lambda$2803 and then computing the EW-weighted mean offset in the velocity space.

The EW-weighted mean velocity of the total absorption profile can be calculated as $v_{\rm tot}$. Assuming the symmetrical component is centered at zero velocity, the mean outflow velocity is given by
\begin{equation}\label{eq:vout}
    v_{\rm out} = \frac{\rm EW_{tot}}{\rm EW_{out}}\cdot v_{\rm tot}.
\end{equation}
The uncertainties in $\rm EW_{out}$ and $v_{\rm out}$ are estimated from our bootstrap analysis.
We apply the aforementioned \texttt{boxcar} measurement to each of the 1000 bootstrap spectra, thereby generating distributions of $\rm EW_{out}$ and $v_{\rm out}$.
The $1\sigma$ uncertainty for each parameter is then quoted as the standard deviation of its respective distribution.

The \texttt{boxcar} method, while model-independent, can yield non-physical results under specific conditions.
In the typical observed outflows, one expects $\rm EW_{out}>0$ and $v_{\rm out}>0$.
However, in the presence of significant noise or complex kinematics, the method may produce $\rm EW_{out}<0$ or $v_{\rm out}<0$.
The robustness of the signal can be evaluated through our bootstrapping analysis.
We note that if the 1$\sigma$ lower limit of $\rm EW_{out}$ is less than zero, the outflow is not statistically significant.
In such cases, according to Equation \ref{eq:vout}, the derived $v_{\rm out}$ becomes unstable to and may diverge to spuriously large values.
We note that the robust detection of $\rm EW_{out}<0$ indicates more absorption in the red wing, which is suggestive of inflow features.
However, this method tailored for outflow characterization is not suitable for the corresponding analysis.

To validate the robustness of our stacking analysis, we also implement the same \texttt{boxcar} measurement on individual galaxy spectra.
First we identify potential outflow candidates by conducting a visual inspection of all JWST spectra which covers \ion{Mg}{2} region, selecting those with discernible \ion{Mg}{2} absorption features.
For these candidates, we calculate the \ion{Mg}{2} absorption EW by integrating the absorbed flux between $2790\ \rm\AA$ and $2808\ \rm\AA$.
The uncertainty in the EW, $\rm EW_{\rm err}$, is derived from the error of the flux densities.
We then apply a S/N cut of $\rm EW_{\rm err}>5$, which yields a final sample of 42 individual JWST spectra with significant \ion{Mg}{2} absorption. The properties and spectra of these sources are tabulated and presented in the Appendix \ref{sec:inds}.

\section{Results}\label{sec:results}

\subsection{Absorption Feature Across Cosmic Time} \label{sec31}

\begin{figure*}[ht!]
\includegraphics[width=1\linewidth]{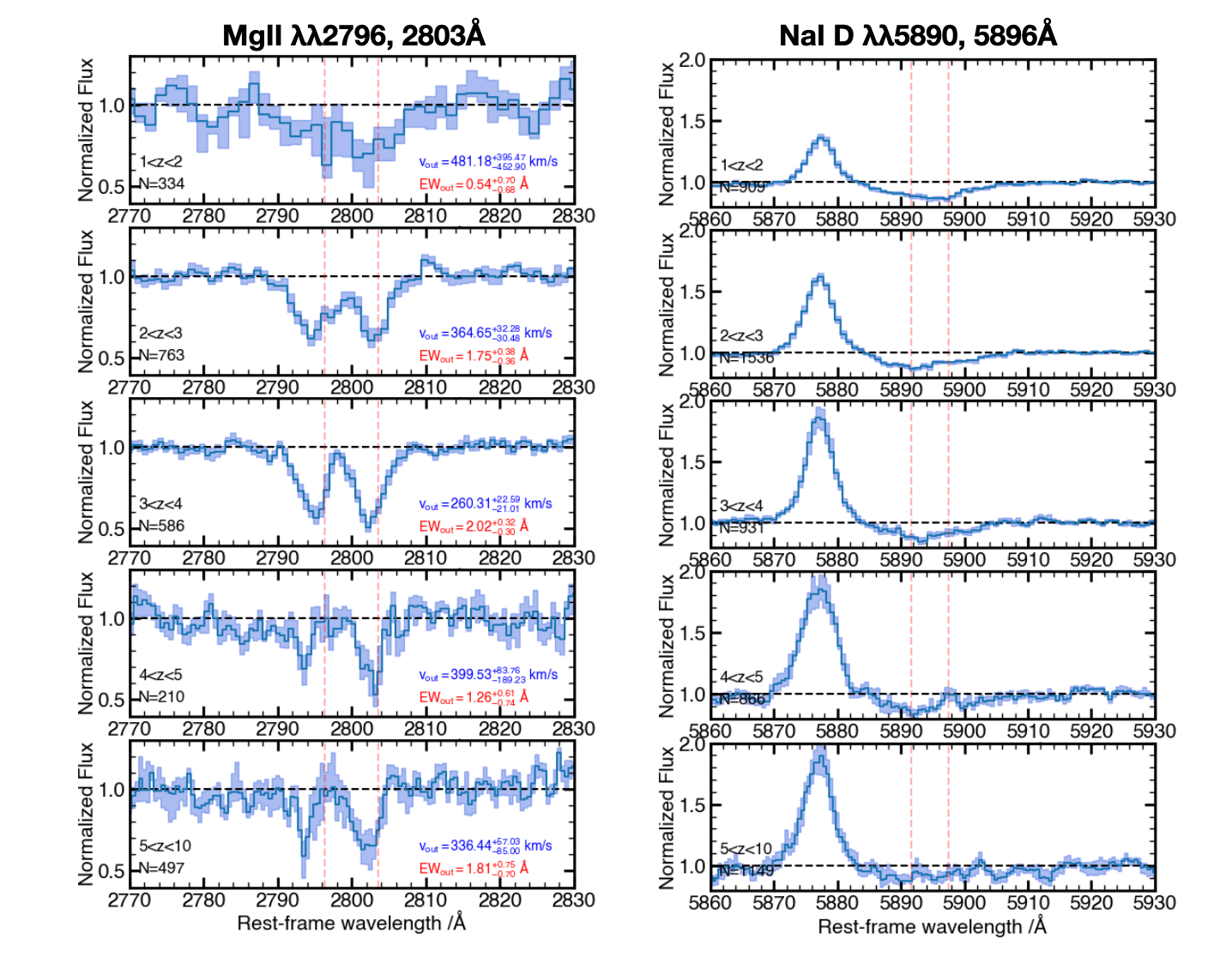}
\caption{Stacked absorption feature in redshift bins. 
The left and right columns show the continuum-normalized stacked spectra centered on the \ion{Mg}{2} $\lambda\lambda2796, 2803$ and \ion{Na}{1}~D $\lambda\lambda5890, 5896$ doublets, respectively. Each row corresponds to a different redshift bin, as indicated in the bottom-left corner of each panel. The blue solid line represents the weighted median spectrum, while the shaded region shows the 1$\sigma$ uncertainty derived from bootstrap resampling. The horizontal dashed line is the continuum level at unity, and the vertical red dashed lines mark the rest-frame wavelengths of the doublet components. The number of spectra in each stack ($N$) and the derived outflow velocity and EW are also annotated.
\label{fig:stacked_z_bin}}
\end{figure*}

Figure~\ref{fig:stacked_z_bin} shows the final stacked spectra for five distinct redshift bins. Our S/N-weighted stacking technique successfully reveals clear absorption features for both the \ion{Mg}{2} and \ion{Na}{1}~D doublets across the cosmic timeline probed by our sample ($1 < z < 10$). For \ion{Mg}{2} doublets, 
despite showing greater spectral fluctuations, the stacked spectra produced using equal-weight stacking exhibit consistent blueshifted absorption features relative to the systemic velocities (marked by the red dashed lines in the left panels of Figure \ref{fig:nw}).
This consistency confirms that the detected outflow signal is robust and not an artifact of our weighting scheme.
The S/N of the composite spectra varies across redshift bins, a difference primarily attributed to the number and intrinsic quality of the spectra contributing to each stack. 
We note that here we focus on establishing the detection of these blueshifted outflows across cosmic time. A robust quantification of their evolutionary trend is not feasible with the current dataset, as the underlying distributions of key galaxy properties (e.g., mass, star formation rate) differ between redshift bins.

The \ion{Na}{1}~D doublet, shown in the right panels, is also detected in the stacked spectra. However, unlike the Mg~II doublet, the \ion{Na}{1}~D profiles do not exhibit clear blueshifted features. As shown in Figure~\ref{fig:stacked_z_bin}, the blue wing of the absorption (where outflow signatures are expected) is severely blended with the strong \ion{He}{1} $\lambda$5876 emission line, hindering both local continuum subtraction and kinematic characterization. 

More fundamentally, the observed \ion{Na}{1}~D absorption is a composite signal. It originates not only from cool, neutral gas in the ISM---the component tracing the outflow---but also from the stellar photospheres within the host galaxy's stellar population. This stellar absorption component arises predominantly from the atmospheres of cool, evolved stars (e.g., K and M giants and supergiants), which can contribute substantially to the total observed EW \citep[e.g.,][]{Heckman2000, Rupke2005}. Disentangling the ISM contribution from this stellar contamination is highly non-trivial and requires sophisticated full spectral fitting with stellar population synthesis (SPS) models. Such modeling is fraught with uncertainties, especially for high-redshift galaxies where the stellar populations are less well-constrained.

Given these combined challenges of emission line blending and inherent stellar contamination, a robust measurement of the ISM outflow properties from the \ion{Na}{1}~D doublet would be highly model-dependent and subject to large systematic uncertainties. Therefore, we focus our subsequent detailed analysis exclusively on the \ion{Mg}{2} doublet as a cleaner tracer of the cool gas outflows. We do not perform further analysis of the \ion{Na}{1}~D feature in stellar mass bins or investigate its potential scaling relations with galaxy properties in this work.

\subsection{Absorption Feature Across Stellar Mass Bins}

\begin{figure*}[ht!]
\includegraphics[width=1\linewidth]{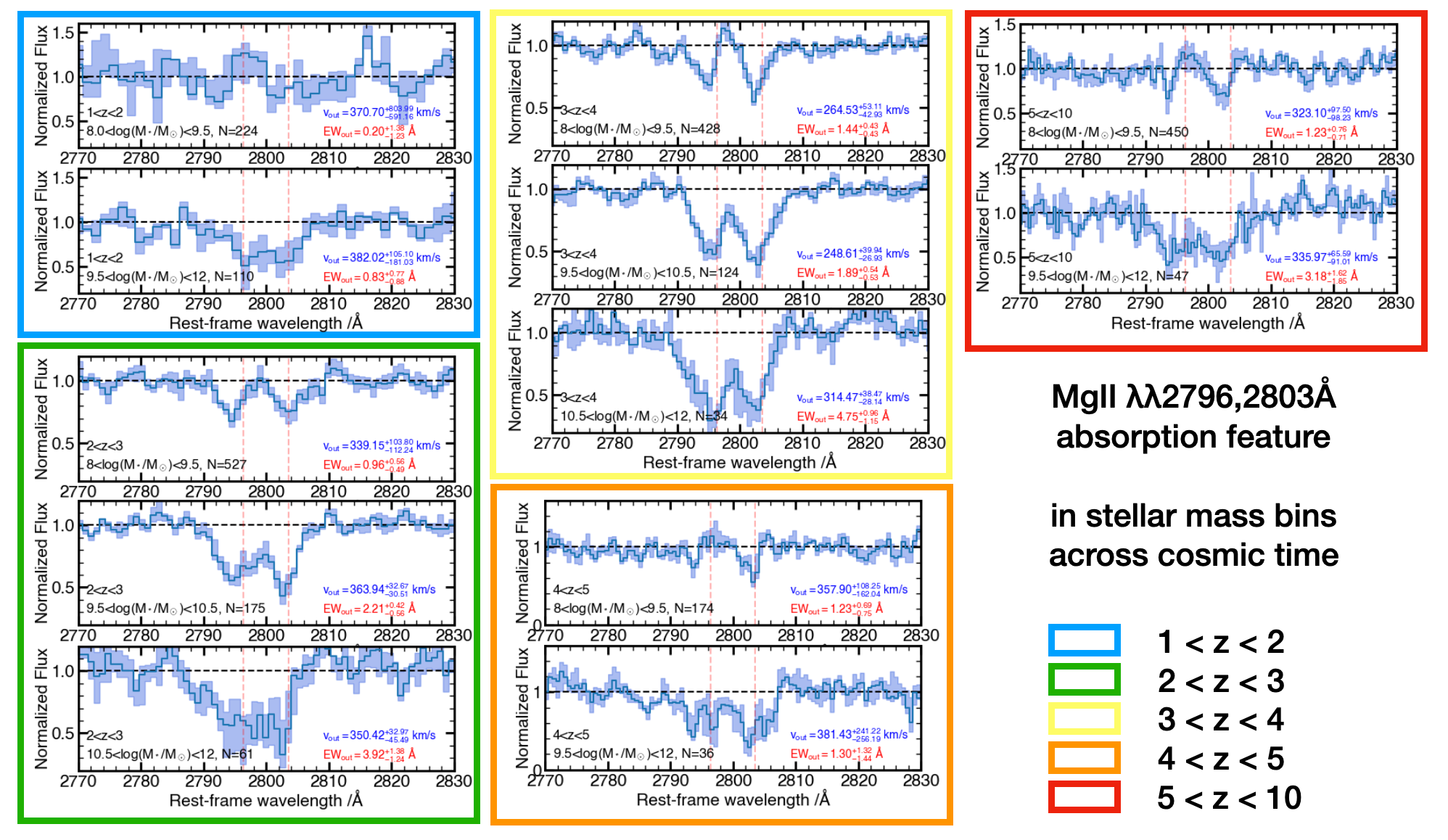}
\caption{Stacked \ion{Mg}{2} feature in stellar mass and redshift bins. 
Each colored box corresponds to a specific redshift bin, as indicated by the legend on the right. Within each box, the panels are arranged in order of increasing stellar mass from top to bottom. The stellar mass range and the number of spectra in each stack are annotated in each panel. The blue solid line, shaded uncertainty region, and dashed lines are the same as in Figure~\ref{fig:stacked_z_bin}.
\label{fig:stacked_z_M_bin}}
\end{figure*}

To explore the dependence of outflow properties on host galaxies, we subdivide our sample into stellar mass bins within each redshift interval. The stellar mass binning is designed to optimize the trade-off between the number of spectra and the S/N ratio, ensuring a robust detection in each bin. The stacked \ion{Mg}{2} absorption features for these bins are presented in Figure~\ref{fig:stacked_z_M_bin}. We find that low-mass bins require a larger number of constituent spectra to achieve a S/N comparable to that of high-mass bins, which could be the result of the relatively low S/N for low-mass galaxy individual spectra.
Blueshifted absorption troughs are detected across all mass bins, confirming the prevalence of cool gas outflows in galaxies of diverse masses.

However, the outflow signal in the $1<z<2$ bins is notably not significant, yielding small $\rm EW_{out}$ values and unstable, spuriously large $v_{\rm out}$ values.
This is likely attributable to the limited sample size and the low spectral S/N of JWST galaxies with \ion{Mg}{2} coverage in this redshift range, despite previous studies having robustly detected outflows at similar epochs \citep{Weiner2009, Bordoloi2014, Yu2025}.
In contrast, the $2<z<3$ and $3<z<4$ bins exhibit high S/N composite spectra, enabling well-constrained measurements of $\rm EW_{out}$ and $v_{\rm out}$ suitable for statistical analysis.
For the highest redshift bins ($4<z<5$ and $5<z<10$), the absorption profiles, though noisy, display clear blueshifts.
This provides strong evidence that cool gas outflows are already present at $z>5$ (corresponding $\sim$ 1 Gyr of the Universe) galaxies. Furthermore, this is the first statistical detection of \ion{Mg}{2}-traced outflow signatures in $z > 3$ galaxies. This finding complements recent studies of ionized outflows \citep[e.g.,][]{Carniani2024, Saldana-Lopez2025, Cooper2025} by confirming that powerful feedback mechanisms are not only present but also effective at entraining the cool gas during the early stages of galaxy evolution.

\begin{figure*}[ht!]
\includegraphics[width=0.5\linewidth]{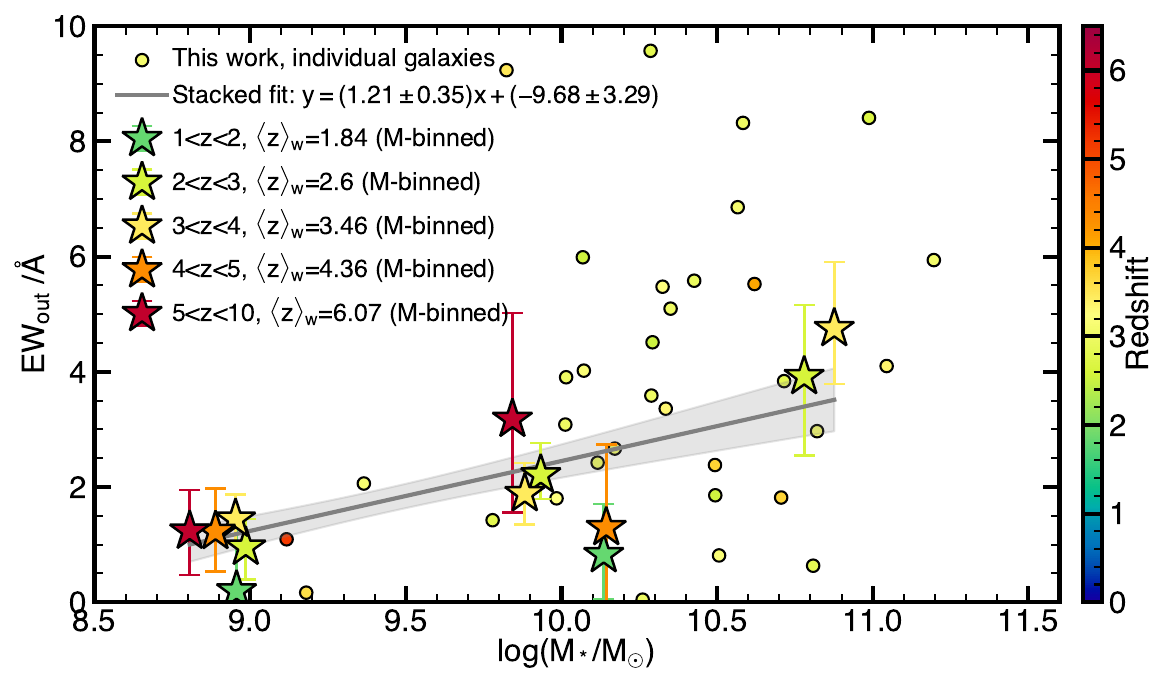}
\includegraphics[width=0.47\linewidth]{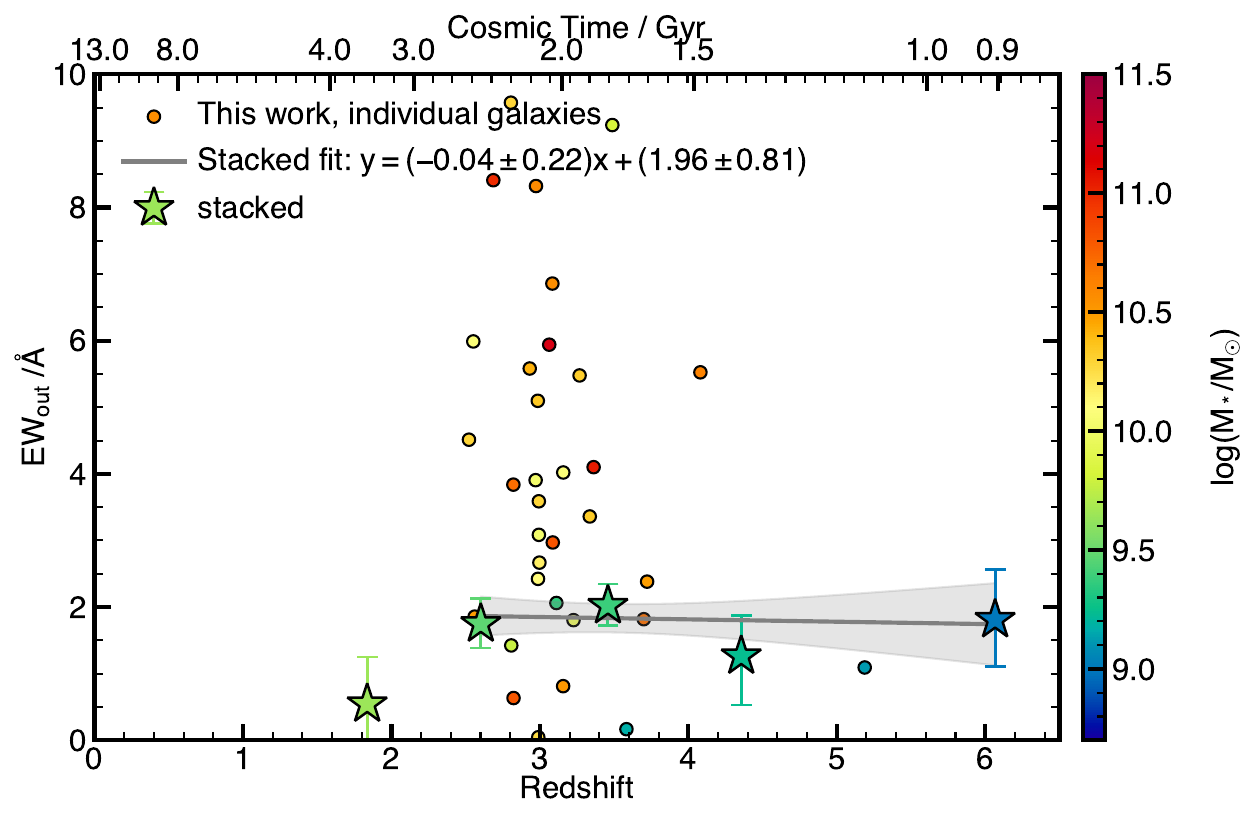}
\caption{Scaling relations for the \ion{Mg}{2} absorption EW. 
\textbf{Left panel:} The outflow EW as a function of stellar mass. Our stacked measurements are shown as star symbols, color-coded by their weighted mean redshift. The grey line marks the best-fit linear relation for the stacked data points, with the shaded region denoting the 1$\sigma$ uncertainty. For comparison, small circles show the measurements from the individual galaxies presented in the Appendix \ref{sec:inds}, color-coded by their redshift.  
\textbf{Right panel:} The outflow EW as a function of redshift, with an additional top axis showing cosmic time. The star symbols again represent our stacked results, but all data points are now color-coded by their weighted mean stellar mass, as indicated by the colorbar.
\label{fig:EW}}
\end{figure*}

\begin{figure*}[ht!]
\includegraphics[width=1\linewidth]{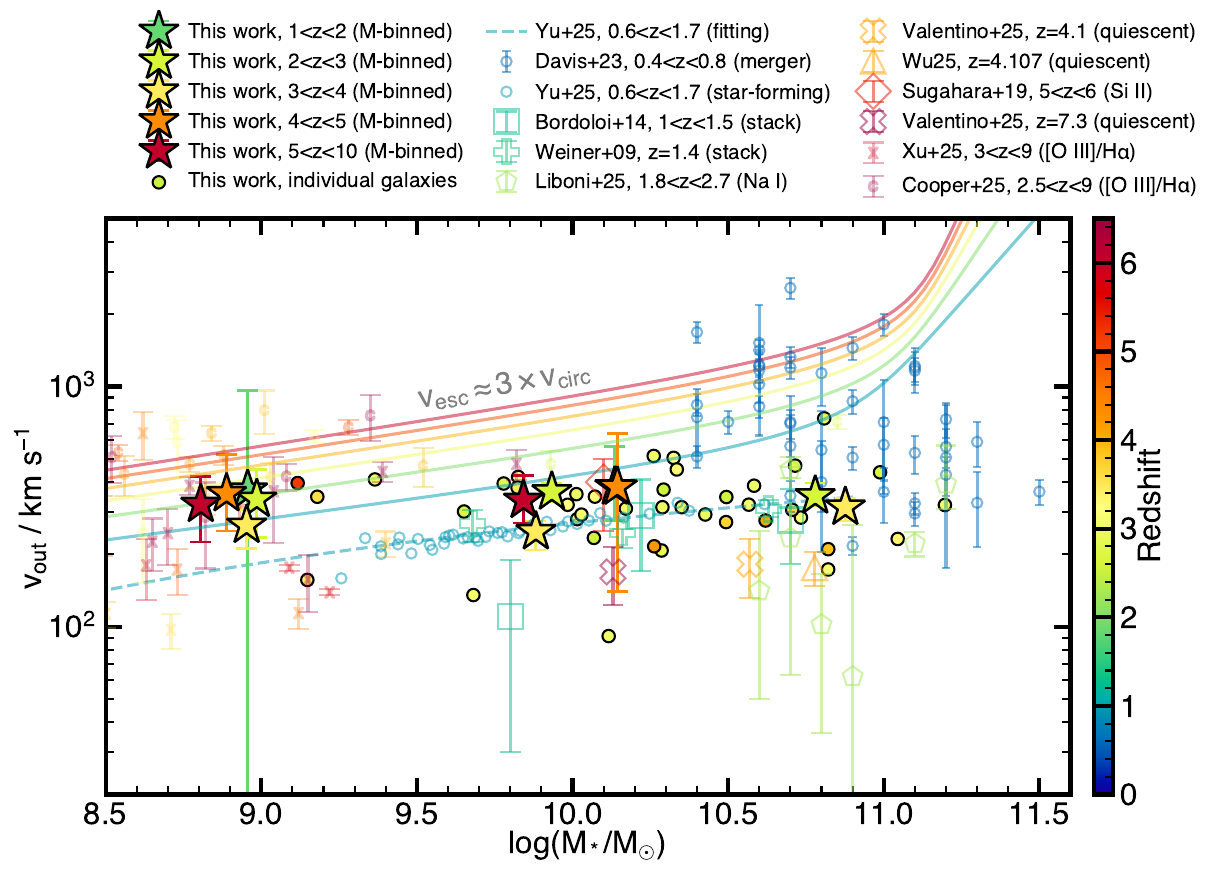}
\caption{The outflow velocity versus stellar mass relation. 
Our measurements from the weighted stacking are shown as star symbols. The color of our data points corresponds to the weighted mean redshift of the stack, as indicated by the color bar on the right. We compare our results with literature data from \citet{Weiner2009}, \citet{Bordoloi2014},\citet{Sugahara2019}, \citet{Davis2023}, \citet{Yu2025}, \citet{Liboni2025}, \citet{Valentino2025},\citet{Cooper2025}, \citet{Wu2025} and \citet{Xu2025}, as detailed in the legend. The colored solid curves represent the estimated escape velocities ($v_{\rm esc} \approx 3 v_{\rm circ}$) as a function of stellar mass at different redshifts ($z=1$ to $z=6$, color-coded), derived using the SHMR of \citet{Moster2013}.}
\label{fig:v_vs_m}
\end{figure*}

\subsection{Scaling Relations of Absorption Equivalent Width}

The scaling relations for the measured \ion{Mg}{2} $\rm EW_{out}$ are displayed in Figure~\ref{fig:EW}.
Our stacked measurements (star symbols) demonstrate that EW$_{\rm out}$ systematically increases with stellar mass across all cosmic epochs. To quantify these trends, we perform a weighted linear least-squares fit to the observed stacked data points $(x_i, y_i \pm \sigma_i)$ using the functional form of $y = a x + b$, where $a$ is the slope, $b$ is the intercept, and $\sigma_i$ denotes the measurement uncertainty (the mean of the upper and lower errors) of the $i$-th data point. The best-fit parameters are estimated by minimizing the weighted residual sum-of-squares $\chi^2 = \sum_i [(y_i - a x_i - b)/\sigma_i]^2$ via the \texttt{scipy.optimize.curve\_fit} algorithm \citep{Virtanen2020}. To ensure the robustness of our results, we exclude the stacked data points at $1<z<2$ from the fitting process (for all scaling relations in this work). For the relation between outflow equivalent width and stellar mass of stacked points (Figure~\ref{fig:EW}, left panel), the derived fitting relation is:
\begin{equation}
\rm EW_{out} = (1.21 \pm 0.35) \log(M_*/M_\odot) + (-9.68 \pm 3.29).
\end{equation}

The measurements for the $z > 3$ bins (including the highest-redshift $5 < z < 10$ bin) are all consistent with the best-fit relation within their $1\sigma$ measurement uncertainties, indicating that galaxies from all epochs broadly follow a universal mass-$\rm EW_{out}$ relation. For comparison, we also plot individual detections from our gallery in Appendix \ref{sec:inds} (small circles), which, despite exhibiting significant scatter at a fixed mass, follow the same average trend.
These sources systematically exhibit larger $\rm EW_{out}$ than the stacked results, indicating that selection effects plays a role in analyzing individual detections.

The right panel of Figure~\ref{fig:EW} shows the outflow EW as a function of redshift. The slope of the redshift relation (in the legend) is statistically consistent with zero, indicating the absence of a simple, monotonic trend with cosmic time. This is primarily a consequence of the strong mass-dependence of the EW, combined with the evolution of the average stellar mass across our stacked samples (as visualized by the color-coding). For instance, the stacked point at $z \sim 6$ has a lower EW than the point at $z \sim 3$, not necessarily because outflows are intrinsically weaker, but because the average mass of the $z \sim 6$ stack is significantly lower. Taken together, these two panels strongly indicate that stellar mass of galaxies is linked to the cool gas outflow EW.

\subsection{Outflow Velocities}

Figure~\ref{fig:v_vs_m} presents our measured outflow velocities as a function of host stellar mass, compared with a compilation of literature results. Our measurements (star symbols) significantly extend previous works of cool gas outflows into the low-mass, high-redshift frontier. A key result from this comparison is the remarkable consistency of outflow velocities at $z > 3$. Across our redshift bins from $z \sim 3$ up to $z \sim 6$ (weighted mean redshift of the highest redshift bin), the outflow velocities remain robustly centered around $\sim 300-400\,\mathrm{km\,s^{-1}}$ and do not exhibit significant evolution.

Cool outflow velocities reported in the literature vary significantly with galaxy properties. 
For instance, vigorous starbursts in merging galaxies can drive outflows with average velocities reaching up to $\sim 1000~\rm km\,s^{-1}$ \citep{Davis2023}. 
In contrast, quiescent galaxies typically exhibit significantly slower winds \citep{Wu2025,Valentino2025}. This difference may stem from the depletion of gas reservoirs and the lack of fresh gas replenishment in quenched systems, whereas typical star-forming galaxies are continuously sustained by gas accretion that fuels energetic feedback. We compare with consistent methodologies from the literature: \citet{Weiner2009} co-added over a thousand spectra and decomposed the \ion{Mg}{2} absorption trough into symmetric and outflow components (we plot the median velocity of their outflow component); \citet{Bordoloi2014} employed a boxcar integration method on stacked \ion{Mg}{2} spectra, which is methodologically similar to this work. 
Our stacked results for typical star-forming galaxies at $z>3$ establish a crucial baseline in this high-redshift regime. While there may be an evolution in outflow properties from $z < 2$ to $z > 2$, our data reveal no evidence for evolution in the scaling relation between outflow velocity and stellar mass at $z>3$.

The mean outflow velocity measured from our individual spectra (mostly at $3<z<4$) are consistent with the scaling relation of \citet{Yu2025}, which was established by stacking low-redshift star-forming main sequence (SFMS) galaxies.
Nonetheless, our stacked results yield systematically higher velocities, particularly at the low-mass end.
Although we have claimed that the measurement in the low-mass bin of $1<z<2$ sample could be affected by numerical instability, the elevated velocities in higher-redshift low-mass bins appear robust. 
This indicates that the high-redshift JWST galaxies in our study may be experiencing more intense, bursty star formation, potentially driving stronger outflows \citep{Looser2025, Clarke2025, Lyu2025}.
This also suggests that the dynamics of the cool gas outflows before the cosmic noon may differ fundamentally from those in the low-redshift universe. 

Crucially, we provide statistical constraints on the velocities of cool gas outflows in low-mass ($M_* \lesssim 10^{9.5}~\rm{M_\odot}$) galaxies at $z > 3$ and towards cosmic dawn. These measurements are vital for constraining feedback models, as they probe the numerous progenitors of today's typical galaxies during their most active growth phases. While there is some evidence for an evolution in the $v_{\rm out}-M_*$ relation from $z<1$ (literature) to $z \approx 2$, we find no significant further evolution at higher redshifts ($z > 3$). We further examine the outflow velocity as a function of cosmic time in Appendix C (Figure \ref{fig:v_vs_z}).

Recently, \citet{Glazer2025} reported a small velocity offset for low-ionization species (LIS) absorption ($v_{\rm cen, LIS} = -23.5 \pm 50.7\,\rm km\,s^{-1}$) from their unweighted mean stack of $z\sim7$ galaxies, in tension with our measured outflow velocities at similar redshifts. However, this discrepancy primarily stems from fundamental differences in methodology. Our work uses a S/N-weighted stack and a `boxcar' method to specifically isolate the high-velocity outflow component. In contrast, \citet{Glazer2025} measured the velocity centroid of a composite LIS profile, created by averaging five transitions (Si\,{\sc ii}~$\lambda$1260, O\,{\sc i}~$\lambda$1302, Si\,{\sc ii}~$\lambda$1304, C\,{\sc ii}~$\lambda$1334, and Si\,{\sc ii}~$\lambda$1526), which reflects the bulk motion of the total cool gas reservoir. Thus, the lower velocity found by \citet{Glazer2025} is a measure of the average gas kinematics and does not contradict our finding of a faster, distinct outflow component. Future work applying a consistent methodology to both datasets will be valuable to harmonize these complementary views of gas kinematics in the early universe.

Combining the results from Figures \ref{fig:EW}, \ref{fig:v_vs_m}, and \ref{fig:v_vs_z}, this lack of significant evolution in the cool outflow properties at $z > 3$ provides a powerful test for theoretical models of galaxy feedback in the early universe. The scaling relations of outflow velocity and equivalent width with stellar mass at $z \sim 3-6$ are roughly consistent with those observed at $z \sim 2$. This suggests that a persistent, unevolving feedback mechanism is already in place at $z > 3$. The strong outflows we detect indicate that stellar feedback remains efficient at driving gas out of galaxies even in the compact, dense environments typical of high-redshift systems. Our findings support a scenario where the interplay between star formation and feedback follows the same physical laws from cosmic noon out to $z \sim 6$, without requiring a fundamental transition in the mode of feedback for the population probed by our sample.

The quantitative measurements of the outflow properties derived from our stacked spectra, including the number of galaxies, weighted-mean redshift and stellar mass, outflow velocities, and EWs for each bin, are fully tabulated in Table~\ref{tab:table}.

\section{Discussion}\label{sec:discussion}

\subsection{The Fate of the Outflowing Gas}

To assess the fate of the outflowing gas, we compare our measured outflow velocities ($v_{\rm out}$) with the escape velocities ($v_{\rm esc}$) of the host dark matter halos. We estimate the halo mass ($M_h$) for each galaxy using the redshift-dependent stellar-to-halo mass relation (SHMR) from \citet{Moster2013}. The virial radius ($R_{\rm vir}$) is then calculated, where the virial overdensity $\Delta_{\rm vir}(z)$ is derived from the approximation of \citet{Bryan1998}. We define the circular velocity of the halo as $v_{\rm circ} \approx \sqrt{GM_h/R_{\rm vir}}$. Following previous studies \citep[e.g.,][]{Veilleux2005, Veilleux2020, Weldon2024}, we adopt a simplified approximation for the escape velocity: $v_{\rm esc} \approx 3 \times v_{\rm circ}$. We caution that these $v_{\rm esc}$ curves are dependent on the assumed redshift evolution of the SHMR, here adopted from \citet{Moster2013}. The SHMR at high redshifts ($z \gtrsim 4$) remains poorly constrained observationally and varies among different theoretical models and simulations \citep[e.g.,][]{Behroozi2019, Girelli2020}, introducing systematic uncertainties to the exact normalization of the escape velocity thresholds.

The resulting $M_* - v_{\rm esc}$ relations for different redshifts are plotted as colored solid curves in Figure \ref{fig:v_vs_m}. As shown in the figure, both our stacked measurements (star symbols) and the majority of individual detections (circles) lie below these escape velocity curves across the entire stellar mass range probed. This indicates that the cool gas outflows traced by \ion{Mg}{2} are predominantly gravitationally bound to their host halos. Instead of escaping into the IGM, this material is likely to decelerate and eventually rain back onto the galaxy, consistent with a ``galactic fountain'' scenario where the baryon cycle is regulated by the recycling of chemically enriched gas. This result aligns with recent statistical studies of ionized gas kinematics using JWST \citep{Cooper2025,Wu2025,Xu2025} that the average outflow velocities are generally insufficient to escape the galaxy potential.

\begin{deluxetable*}{rccccccc}
\tablewidth{0pt}
\tablecaption{Properties of Stacked Samples and Outflow Measurements. The columns are as follows: (1) Redshift bin. (2) Weighted mean redshift of the bin. (3) Stellar mass bin. For rows with a dash, measurements correspond to the entire redshift bin. (4) Weighted mean of the logarithm of the stellar mass. (5) Number of spectra in the stack. (6) Weighted mean outflow velocity derived from the \ion{Mg}{2} doublet. (7) Outflow equivalent width of the \ion{Mg}{2} doublet. (8) Lower limit on mass outflow rate.
}
\label{tab:table}
\tablehead{
\colhead{$z$} &
\colhead{$\langle z\rangle_w$} &
\colhead{$\log M_{*}$} &
\colhead{$\langle\log M_{*}\rangle_w$} &
\colhead{$N$} &
\colhead{$v_{\rm out}$} &
\colhead{$\rm{EW_{\rm out}}$} &
\colhead{$>\dot{M}_\mathrm{out,min}$}
\\
\colhead{}&\colhead{}&\colhead{($\rm M_\odot$)}&\colhead{($\rm M_\odot$)}&\colhead{}&\colhead{($\rm km\,s^{-1}$)}&\colhead{(\AA)}&
\colhead{($\mathrm{M}_{\odot}\,\mathrm{yr}^{-1}$)}
}
\startdata
1--2$^*$ & 1.836 & -- & 9.65 & 334 & $481^{+395}_{-453}$ & $0.54^{+0.70}_{-0.68}$ & $0.089$ \\
1--2$^*$ & 1.828 & 8.0--9.5 & 8.96 & 224 & $371^{+804}_{-591}$ & $0.20^{+1.38}_{-1.23}$ & $0.068$ \\
1--2$^*$ & 1.842 & 9.5--12.0 & 10.14 & 110 & $382^{+105}_{-181}$ & $0.83^{+0.77}_{-0.88}$ & $0.070$ \\
2--3 & 2.602 & -- & 9.47 & 763 & $365^{+32}_{-30}$ & $1.75^{+0.38}_{-0.36}$ & $0.067$ \\
2--3 & 2.608 & 8.0--9.5 & 8.99 & 527 & $339^{+104}_{-112}$ & $0.96^{+0.56}_{-0.49}$ & $0.062$ \\
2--3 & 2.582 & 9.5--10.5 & 9.93 & 175 & $364^{+33}_{-31}$ & $2.21^{+0.42}_{-0.56}$ & $0.067$ \\
2--3 & 2.648 & 10.5--12.0 & 10.78 & 61 & $350^{+33}_{-45}$ & $3.92^{+1.38}_{-1.24}$ & $0.065$ \\
3--4 & 3.458 & -- & 9.39 & 586 & $260^{+23}_{-21}$ & $2.02^{+0.32}_{-0.30}$ & $0.048$ \\
3--4 & 3.486 & 8.0--9.5 & 8.95 & 428 & $265^{+53}_{-43}$ & $1.44^{+0.43}_{-0.43}$ & $0.049$ \\
3--4 & 3.410 & 9.5--10.5 & 9.88 & 124 & $249^{+40}_{-27}$ & $1.89^{+0.54}_{-0.53}$ & $0.046$ \\
3--4 & 3.423 & 10.5--12.0 & 10.88 & 34 & $314^{+38}_{-28}$ & $4.75^{+0.96}_{-1.15}$ & $0.058$ \\
4--5 & 4.356 & -- & 9.24 & 210 & $400^{+84}_{-189}$ & $1.26^{+0.61}_{-0.74}$ & $0.074$ \\
4--5 & 4.385 & 8.0--9.5 & 8.89 & 174 & $358^{+108}_{-162}$ & $1.23^{+0.69}_{-0.75}$ & $0.066$ \\
4--5 & 4.281 & 9.5--12.0 & 10.14 & 36 & $381^{+241}_{-256}$ & $1.30^{+1.32}_{-1.44}$ & $0.070$ \\
5--10 & 6.067 & -- & 9.01 & 497 & $336^{+57}_{-85}$ & $1.81^{+0.75}_{-0.70}$ & $0.062$ \\
5--10 & 6.078 & 8.0--9.5 & 8.81 & 450 & $323^{+98}_{-98}$ & $1.23^{+0.76}_{-0.71}$ & $0.060$ \\
5--10 & 6.017 & 9.5--12.0 & 9.84 & 47 & $336^{+66}_{-91}$ & $3.18^{+1.62}_{-1.85}$ & $0.062$ \\
\enddata
\tablecomments{The quoted uncertainties on $v_{\rm out}$ and EW represent the 1$\sigma$ confidence intervals derived from 1000 bootstrap resampling iterations. The outflow properties are measured using the \texttt{boxcar} method described in Section \ref{sec:method}. *: The velocity measurement is considered unconstrained due to the low significance of the $\rm EW_{out}$ detection ($1\sigma$ lower limit of $\rm EW_{out}$ is less than zero, see Section \ref{sec:method}).}
\end{deluxetable*}

\subsection{Lower Limits on Mass Outflow Rates}\label{sec:mdot}

While deriving precise mass outflow rates is challenging due to uncertainties in the ionization structure and outflow geometry, we can estimate robust, conservative lower limits to facilitate comparisons with theoretical models. Given that the observed \ion{Mg}{2} absorption profiles are generally saturated (doublet ratio $\sim 1$), we adopt a minimum column density of $N(\text{\ion{Mg}{2}}) \gtrsim 10^{14}$ cm$^{-2}$. To derive a strict lower limit on the total gas mass, we assume a solar abundance ratio \citep[$\log(\text{Mg/H})_{\odot} \approx -4.4$;][]{Asplund2009} and apply no ionization correction (i.e., assuming the gas is entirely in the \ion{Mg}{2} state). Furthermore, we adopt a minimal outflow radius of $R_{\text{out}} = 1$ kpc, which is significantly more compact than the typical circumgalactic medium scales ($\sim 5-10$ kpc) assumed in other works.

Based on the conservative assumptions of a minimal launch radius ($r_{\rm min}=1$~kpc) and a saturated column density limit ($N_{\rm H, min} \approx 2.5 \times 10^{18} \, {\rm cm}^{-2}$), the mass outflow rate lower limit is estimated by scaling the standard relation \citep{Weiner2009}:
\begin{equation}
    \dot{M}_{\rm out, min} = 22 \, {\rm M}_{\odot} \, {\rm yr}^{-1} \cdot C_f \cdot \frac{N_{\rm H, min}}{10^{20} \, {\rm cm}^{-2}} \cdot \frac{1\,{\rm kpc}}{5\,{\rm kpc}} \cdot \frac{v_{\rm out}}{300 \, {\rm km \, s}^{-1}},
\end{equation}
where $C_f \approx 0.5$\citep{Pessa2024} is the covering fraction and $N_{\rm H, min}$ is derived from $N({\rm Mg\,II}) > 10^{14}\,{\rm cm}^{-2}$ assuming solar abundance. These derived lower limits for stackted results are listed in Table~\ref{tab:table}. We note that these values are strictly minimal constraints; realistic mass outflow rates, accounting for larger radii ($R \gtrsim 5$ kpc) and ionization corrections (\ion{Mg}{2} fraction $f \lesssim 0.1$), are likely an order of magnitude higher.

\subsection{Synergy of Stacking and Individual Analysis}

A key strength of our work is the synergistic combination of a statistical stacking analysis with a targeted study of individual detections.
As illustrated in Figure~\ref{fig:appendix_individuals} (Appendix \ref{sec:inds}) , the galaxies with the most prominent, high-S/N absorption features are typically massive ($M_* > 10^{10}~\rm{M_\odot}$). 
These individual detections serve as unambiguous proof-of-concept, confirming the physical reality of the outflow phenomenon and revealing the rich diversity of absorption profile morphologies. This diversity includes not only powerful outflows but also a few intriguing cases of redshifted absorption as potential signatures of gas accretion (inflow), which we defer to a future, more detailed analysis (see Appendix \ref{sec:inds}).

However, these massive objects represent only a small fraction of the galaxy population. By extending our analysis to lower stellar masses ($M_* < 10^{9.5} M_\odot$) via spectral stacking, we demonstrate that cool gas outflows are not limited to rare, massive starbursts but are a ubiquitous feature of typical galaxies at these epochs. The combination of individual characterization and statistical stacking thus provides a complete picture of feedback across the full dynamic range of our sample.

\section{Summary}\label{sec:summary}

We present the first statistical study of \ion{Mg}{2}-traced cool gas outflows in galaxies spanning a wide cosmic timeline from $z \approx 1$ to $z > 6$ (with sample coverage extending to $z \sim 10$). Leveraging thousands of public JWST/NIRSpec spectra, we employ a S/N-weighted spectral stacking technique, complemented by an analysis of individual high-S/N detections, to characterize the faint absorption signatures of the \ion{Mg}{2} doublet. Our sample is shown to be broadly representative of the underlying galaxy population. Our main findings are summarized as follows:

\begin{enumerate}
    \item We robustly detect blueshifted \ion{Mg}{2} absorption, indicative of cool galactic outflows, in nearly all redshift and stellar mass bins from $z \approx 1$ to $z > 6$, with the exception of the lowest-mass bin ($\log M_*/M_\odot < 9.5$) at $z < 2$ where the signal is marginal. The derived typical outflow velocities ($\sim 300-400$\,km\,s$^{-1}$) suggest that cool gas outflows are a common feature of galaxies across the probed cosmic history.

    \item The properties of the cool gas outflows appear correlated with the stellar mass of the host galaxy. We observe that the outflow equivalent width (EW$_{\rm out}$) systematically increases with stellar mass at all cosmic epochs (with the fitted slope of $1.21 \pm 0.35$). This trend suggests that more massive galaxies are associated with stronger absorption from cool, outflowing gas reservoirs.
    
    \item We find no evidence for significant evolution in the outflow scaling relations at high redshifts ($z > 3$). Both the relation between EW$_{\rm out}$ and stellar mass, as well as the outflow velocities, are broadly consistent from $z \approx 3$ to $z \approx 6$ (where our highest redshift bin is centered). 
    These consistencies suggest that the feedback mechanisms governing the cool gas phase operate efficiently even at the epoch of cosmic dawn.

    \item The average outflow velocities measured in our stacks generally remain below the estimated escape velocities of the host dark matter halos. This indicates that the bulk of the cool gas traced by \ion{Mg}{2} is likely gravitationally bound and may eventually decelerate and return to the galaxy. These results are consistent with a ``galactic fountain'' scenario, where the baryon cycle is regulated by the recycling of chemically enriched gas rather than by efficient removal into the intergalactic medium.
\end{enumerate}

Our results provide fundamental, statistically observational constraints on the interplay between galactic feedback and the baryon cycle over 12 billion years of cosmic history. By characterizing outflow scaling relations in the high-redshift ($z > 3$), low-mass regime, we provide statistical constraints on the cool gas phase that were previously lacking, complementing recent studies of ionized outflows \citep[e.g.,][]{Carniani2024, Cooper2025, Xu2025}. These findings will be crucial for calibrating the next generation of cosmological simulations and refining our theoretical understanding of galaxy evolution towards cosmic dawn.

\begin{acknowledgments}
The authors thank the anonymous referee for the helpful comments that improved the quality of this Letter. EW thanks support of the National Science Foundation of China (Nos. 12473008). CL is supported by the Fundamental Research Funds for the Central Universities (WK2030250123). The authors gratefully acknowledge the support of Cyrus Chun Ying Tang Foundations.

The data products presented herein are retrieved from the Dawn JWST Archive (DJA). DJA is an initiative of the Cosmic Dawn Center (DAWN), which is funded by the Danish National Research Foundation under grant DNRF140.
\end{acknowledgments}

\appendix
\section{Comparison with Equal-Weight Stacking}
\label{sec:app}

\begin{figure*}[ht!]
\includegraphics[width=1\linewidth]{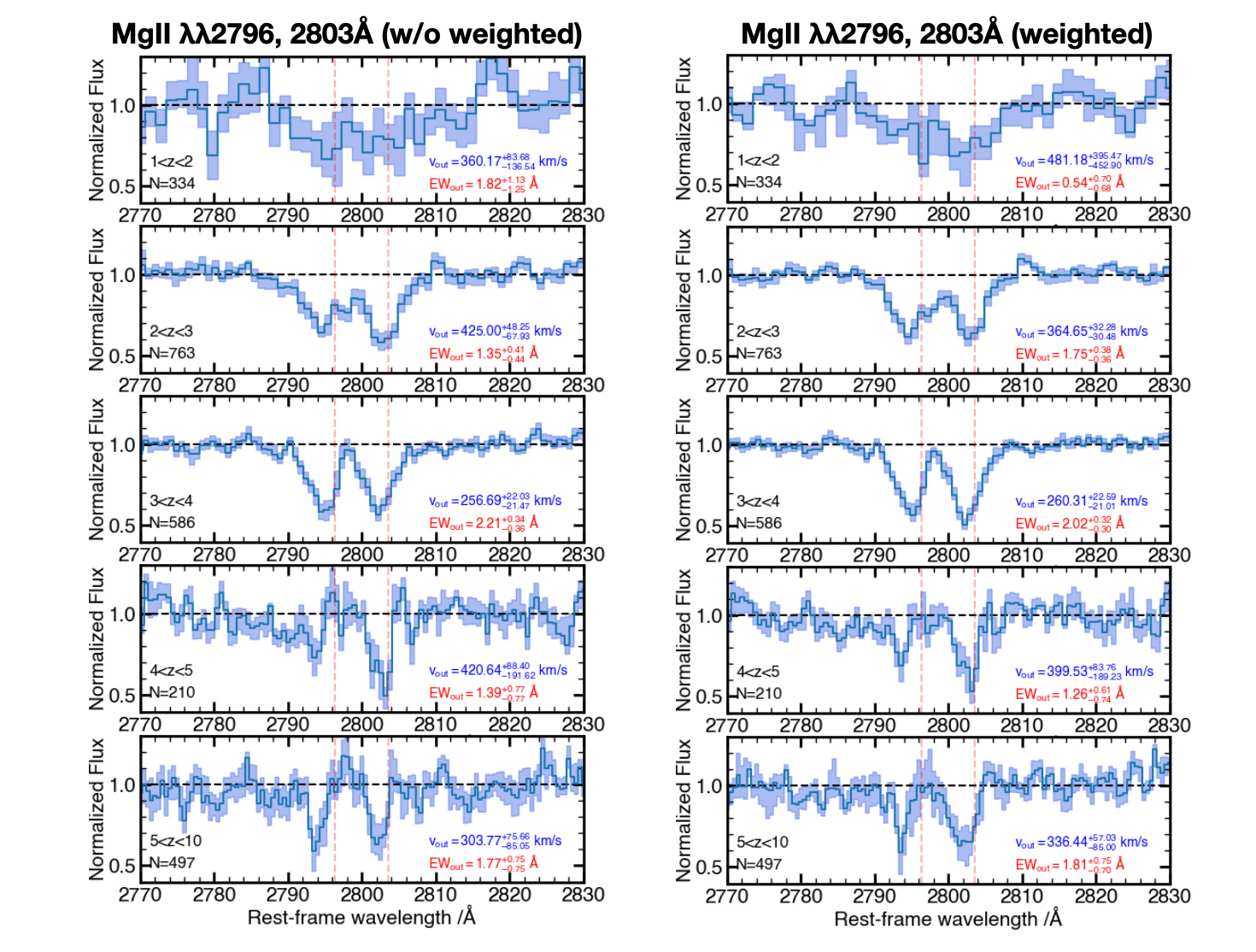}
\caption{Comparison with equal-weight stacking. The left panels show the resulting stacked \ion{Mg}{2} profiles for each redshift bin when using an unweighted (equal-weight) median stacking method. These profiles should be directly compared with the results from our fiducial S/N-weighted method (right panels, the same as in the left panels in Figure \ref{fig:stacked_z_bin}). Note the significantly lower S/N and less-defined absorption features compared to the weighted stacks, which demonstrates the superiority of the fiducial method.
\label{fig:nw}}
\end{figure*}

In addition to the S/N-weighted stacking method adopted in our main analysis (Section~\ref{sec:method}), we perform a comparative test using an unweighted (equal-weight) stacking approach. In this alternative scheme, every spectrum contributes equally to the final stack, regardless of its individual S/N.

The left panels of Figure~\ref{fig:nw} presents the results of this test. A direct visual comparison with our fiducial stacks (right panels, the same as the left panels of Figure~\ref{fig:stacked_z_bin}) immediately reveals the inferiority of this simpler method. 
Through involving poorly normalized individual spectra, composite spectra derived from unweighted stacking exhibit larger noise which is non-physical.
This may potentially wash out the subtle, underlying absorption features. As a result, the continuum is more poorly defined, and the overall statistical significance of the detection is substantially lower. This comparison therefore underscores the critical importance of our S/N-weighting scheme for robustly detecting and characterizing the faint outflow signals in our sample.
Although the effect of weighting is more subtle in our current, relatively small JWST sample, its importance becomes dramatic when stacking larger datasets to recover very faint signals.

\section{Individual Galaxy Spectra of Absorbers}
\label{sec:inds}

\begin{figure*}[ht!]
\includegraphics[width=1\linewidth]{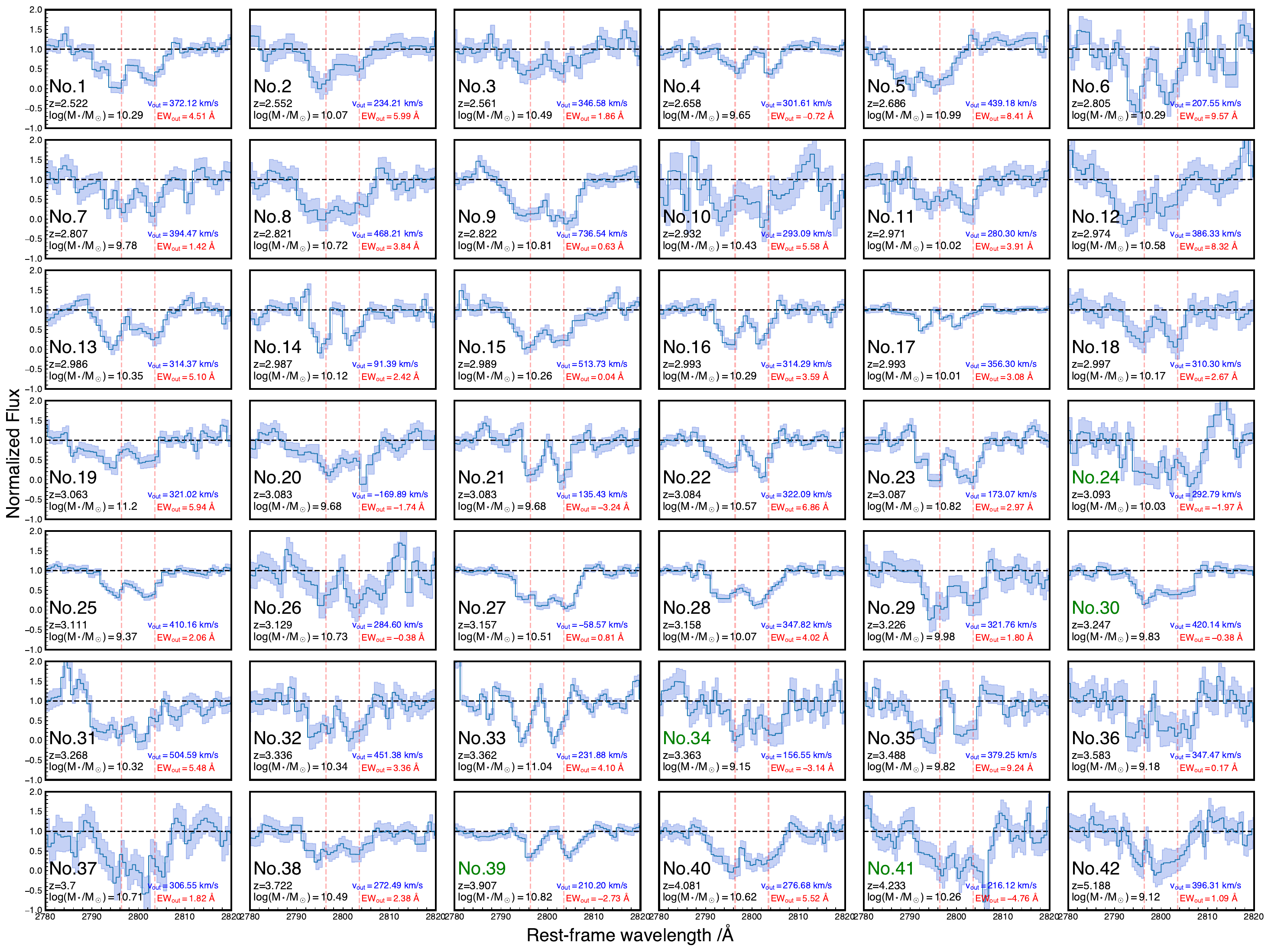}
\caption{Individual \ion{Mg}{2} absorption detections. Each panel shows the continuum-normalized spectrum of an individual galaxy from our sample, centered on the \ion{Mg}{2} doublet region, with the panels arranged in order of increasing redshift. The shaded regions are 1$\sigma$ uncertainties and the dashed lines mark the rest-frame \ion{Mg}{2} doublet. Each spectrum is annotated with its spectroscopic redshift ($z$), stellar mass ($\log(M_*/\rm{M_\odot})$), and the \texttt{boxcar}-measured velocity ($v_{\rm out}$) and equivalent width ($\rm{EW_{out}}$). While most objects show blueshifted absorption indicative of outflows, a few exhibit visually redshifted absorption (marked by green color), which may be a signature of gas inflow. \label{fig:appendix_individuals}}
\end{figure*}

\begin{deluxetable*}{rccccccc} 
\tablewidth{0pt}
\tablecaption{Properties of Individual Galaxy with \ion{Mg}{2} Absorption Detections. The columns are as follows: (1) Object number. (2) Object name in the DJA catalog. (3) Right Ascension. (4) Declination. (5) Spectroscopic redshift. (6) Logarithm of the stellar mass. (7) Mean outflow velocity derived from the \ion{Mg}{2} doublet. (8) Outflow equivalent width of the \ion{Mg}{2} doublet. 
}
\label{tab:table_inds}
\tablehead{
\colhead{No.} &
\colhead{Name in DJA catalog} &
\colhead{R.A.} &
\colhead{Dec.} &
\colhead{$z_{\rm{spec}}$} &
\colhead{$\log M_{*}$} &
\colhead{$v_{\rm out}$} &
\colhead{$\rm{EW_{out}}$} 
\\
\colhead{} &
\colhead{} &
\colhead{(deg)} &
\colhead{(deg)} &
\colhead{} &
\colhead{($\rm{M}_{\odot}$)} &
\colhead{($\rm{km\ s^{-1}}$)} &
\colhead{(\AA)}
}
\startdata
1 & cosmos-curti-v4\_g140m-f100lp\_1879\_3747 & $150.06155$ & $2.23362$ & 2.522 & 10.29 & $372^{+40}_{-39}$ & $4.51^{+1.02}_{-1.02}$ \\
2 & bluejay-north-v4\_g140m-f100lp\_1810\_16474 & $150.11344$ & $2.34747$ & 2.552 & 10.07 & $234^{+80}_{-58}$ & $5.99^{+1.82}_{-1.75}$ \\
3 & bluejay-south-v4\_g140m-f100lp\_1810\_8888 & $150.12873$ & $2.26979$ & 2.561 & 10.49 & $347^{+199}_{-179}$ & $1.86^{+1.92}_{-2.01}$ \\
4 & cosmos-curti-v4\_g140m-f100lp\_1879\_1858 & $150.08136$ & $2.20749$ & 2.658 & 9.65 & $302^{+276}_{-311}$ & $-0.72^{+0.94}_{-0.94}$ \\
5 & bluejay-south-v4\_g140m-f100lp\_1810\_8002 & $150.11993$ & $2.26102$ & 2.686 & 10.99 & $439^{+21}_{-22}$ & $8.41^{+1.21}_{-1.13}$ \\
6 & jades-gds03-v4\_g140m-f070lp\_1286\_181082 & $53.12218$ & $-27.85834$ & 2.805 & 10.29 & $208^{+58}_{-42}$ & $9.57^{+2.45}_{-2.30}$ \\
7 & jades-gds1-v4\_g140m-f070lp\_1286\_42789 & $53.18607$ & $-27.80225$ & 2.807 & 9.78 & $394^{+247}_{-358}$ & $1.42^{+1.95}_{-1.72}$ \\
8 & gds-udeep-v4\_g140m-f070lp\_3215\_199773 & $53.16324$ & $-27.80906$ & 2.821 & 10.72 & $468^{+67}_{-144}$ & $3.84^{+1.80}_{-1.96}$ \\
9 & bluejay-north-v4\_g140m-f100lp\_1810\_18071 & $150.13527$ & $2.36361$ & 2.822 & 10.81 & $737^{+1296}_{-955}$ & $0.63^{+1.24}_{-1.19}$ \\
10 & jades-gdn2-v4\_g140m-f070lp\_1181\_20185 & $189.10652$ & $62.23063$ & 2.932 & 10.43 & $293^{+115}_{-117}$ & $5.58^{+2.90}_{-2.98}$ \\
11 & jades-gds04-v4\_g140m-f070lp\_1286\_53237 & $53.04188$ & $-27.86134$ & 2.971 & 10.02 & $280^{+152}_{-86}$ & $3.91^{+1.91}_{-1.87}$ \\
12 & jades-gdn10-v4\_g140m-f070lp\_1181\_77071 & $189.22890$ & $62.19548$ & 2.974 & 10.58 & $386^{+39}_{-53}$ & $8.32^{+2.36}_{-2.26}$ \\
13 & bluejay-north-v4\_g140m-f100lp\_1810\_17793 & $150.13779$ & $2.36100$ & 2.986 & 10.35 & $314^{+40}_{-36}$ & $5.10^{+0.95}_{-1.12}$ \\
14 & bluejay-north-v4\_g140m-f100lp\_1810\_17906 & $150.12949$ & $2.36170$ & 2.987 & 10.12 & $91^{+159}_{-88}$ & $2.42^{+1.06}_{-0.99}$ \\
15 & jades-gdn11-v4\_g140m-f070lp\_1181\_25603 & $189.17021$ & $62.23287$ & 2.989 & 10.26 & $514^{+1683}_{-1014}$ & $0.04^{+0.87}_{-0.90}$ \\
16 & aurora-gdn02-v4\_g140m-f100lp\_1914\_22384 & $189.16149$ & $62.25202$ & 2.993 & 10.29 & $314^{+56}_{-46}$ & $3.59^{+0.96}_{-0.89}$ \\
17 & aurora-gdn02-v4\_g140m-f100lp\_1914\_19848 & $189.16046$ & $62.23943$ & 2.993 & 10.01 & $356^{+18}_{-21}$ & $3.08^{+0.39}_{-0.41}$ \\
18 & jades-gds-wide-v4\_g140m-f070lp\_1180\_11929 & $53.13048$ & $-27.77596$ & 2.997 & 10.17 & $310^{+182}_{-153}$ & $2.67^{+2.16}_{-2.06}$ \\
19 & jades-gds-wide3-v4\_g140m-f070lp\_1180\_197911 & $53.16531$ & $-27.81414$ & 3.063 & 11.20 & $321^{+32}_{-30}$ & $5.94^{+1.13}_{-1.13}$ \\
20 & gds-rieke-v4\_g140m-f100lp\_1207\_54612 & $53.14468$ & $-27.77118$ & 3.083 & 9.68 & $-170^{+824}_{-313}$ & $-1.74^{+1.52}_{-1.47}$ \\
21 & jades-gds1-v4\_g140m-f070lp\_1286\_54612 & $53.14468$ & $-27.77118$ & 3.083 & 9.68 & $135^{+176}_{-92}$ & $-3.24^{+1.27}_{-1.36}$ \\
22 & aurora-gdn01-v4\_g140m-f100lp\_1914\_4113 & $150.13793$ & $2.22025$ & 3.084 & 10.57 & $322^{+21}_{-20}$ & $6.86^{+0.92}_{-0.91}$ \\
23 & cosmos-curti-v4\_g140m-f100lp\_1879\_599 & $150.09334$ & $2.18999$ & 3.087 & 10.82 & $173^{+141}_{-78}$ & $2.97^{+1.22}_{-1.11}$ \\
24 & bluejay-south-v4\_g140m-f100lp\_1810\_7934 & $150.09731$ & $2.26012$ & 3.093 & 10.03 & $293^{+249}_{-186}$ & $-1.97^{+2.18}_{-2.48}$ \\
25 & aurora-gdn02-v4\_g140m-f100lp\_1914\_21033 & $189.21160$ & $62.24571$ & 3.111 & 9.37 & $410^{+65}_{-62}$ & $2.06^{+0.61}_{-0.67}$ \\
26 & jades-gdn11-v4\_g140m-f070lp\_1181\_25351 & $189.17337$ & $62.23041$ & 3.129 & 10.73 & $285^{+336}_{-280}$ & $-0.38^{+2.61}_{-2.64}$ \\
27 & aurora-gdn01-v4\_g140m-f100lp\_1914\_4740 & $150.15878$ & $2.22606$ & 3.157 & 10.51 & $-59^{+524}_{-206}$ & $0.81^{+0.58}_{-0.62}$ \\
28 & aurora-gdn01-v4\_g140m-f100lp\_1914\_6143 & $150.18320$ & $2.24024$ & 3.158 & 10.07 & $348^{+30}_{-28}$ & $4.02^{+0.65}_{-0.72}$ \\
29 & excels-uds04-v4\_g140m-f100lp\_3543\_61762 & $34.27260$ & $-5.22469$ & 3.227 & 9.98 & $322^{+220}_{-244}$ & $1.80^{+2.02}_{-2.05}$ \\
30 & aurora-gdn01-v4\_g140m-f100lp\_1914\_8363 & $150.13641$ & $2.26468$ & 3.247 & 9.83 & $420^{+623}_{-359}$ & $-0.38^{+0.82}_{-0.79}$ \\
31 & bluejay-south-v4\_g140m-f100lp\_1810\_8338 & $150.07943$ & $2.26452$ & 3.268 & 10.32 & $505^{+62}_{-82}$ & $5.48^{+1.62}_{-1.43}$ \\
32 & jades-gdn2-v4\_g140m-f070lp\_1181\_2945 & $189.14089$ & $62.28066$ & 3.336 & 10.34 & $451^{+100}_{-176}$ & $3.36^{+1.99}_{-1.94}$ \\
33 & aurora-gdn02-v4\_g140m-f100lp\_1914\_11584 & $189.10502$ & $62.28033$ & 3.362 & 11.04 & $232^{+47}_{-40}$ & $4.10^{+0.89}_{-0.90}$ \\
34 & aurora-gdn02-v4\_g140m-f100lp\_1914\_23927 & $189.24105$ & $62.25951$ & 3.363 & 9.15 & $157^{+249}_{-121}$ & $-3.14^{+1.92}_{-2.08}$ \\
35 & bluejay-south-v4\_g140m-f100lp\_1810\_13297 & $150.11868$ & $2.31445$ & 3.488 & 9.82 & $379^{+30}_{-34}$ & $9.24^{+1.68}_{-1.68}$ \\
36 & jades-gds03-v4\_g140m-f070lp\_1286\_180657 & $53.11252$ & $-27.85900$ & 3.583 & 9.18 & $347^{+327}_{-357}$ & $0.17^{+2.41}_{-2.54}$ \\
37 & excels-uds04-v4\_g140m-f100lp\_3543\_39063 & $34.29044$ & $-5.26208$ & 3.700 & 10.71 & $307^{+306}_{-292}$ & $1.82^{+3.04}_{-2.94}$ \\
38 & excels-uds04-v4\_g140m-f100lp\_3543\_56133 & $34.30228$ & $-5.23394$ & 3.722 & 10.49 & $272^{+108}_{-78}$ & $2.38^{+1.26}_{-1.22}$ \\
39 & jades-gdn10-v4\_g140m-f070lp\_1181\_82948 & $189.23303$ & $62.23678$ & 3.907 & 10.82 & $210^{+80}_{-50}$ & $-2.73^{+0.66}_{-0.73}$ \\
40 & excels-uds02-v4\_g140m-f100lp\_3543\_117004 & $34.28827$ & $-5.13754$ & 4.081 & 10.62 & $277^{+35}_{-37}$ & $5.52^{+1.18}_{-1.23}$ \\
41 & excels-uds03-v4\_g140m-f100lp\_3543\_45052 & $34.36713$ & $-5.25202$ & 4.233 & 10.26 & $216^{+138}_{-83}$ & $-4.76^{+2.13}_{-2.04}$ \\
42 & aurora-gdn02-v4\_g140m-f100lp\_1914\_100067 & $189.23479$ & $62.25748$ & 5.188 & 9.12 & $396^{+265}_{-297}$ & $1.09^{+1.79}_{-1.65}$ \\
\enddata
\tablecomments{The outflow properties are measured using the \texttt{boxcar} method described in Section \ref{sec:method}.\\
$^*$Typical outflows correspond to $\rm EW_{out}>0$ and $v_{\rm out}>0$. Any result with $\rm EW_{out}<0$ or $v_{\rm out}<0$ is purely numerical outputs (see Section \ref{sec:method}). Note that these negative values do not necessarily indicate gas inflows, as our measurement methodology is not designed to characterize inflow kinematics.}
\end{deluxetable*}

While our main analysis relies on spectral stacking to overcome the low S/N inherent in most individual high-redshift spectra, a subset of galaxies in our sample do exhibit individually detectable \ion{Mg}{2} absorption features. In Figure~\ref{fig:appendix_individuals}, we present a postage stamp gallery of such individual detections, selected to illustrate the presence of these features in our dataset. The corresponding properties of these individual galaxies and outflow measurements are tabulated in Table \ref{tab:table_inds}.

These examples highlight the significant kinematic diversity present in the underlying population. The majority of the selected objects display clear blueshifted absorption profiles, providing direct visual confirmation of the outflow signatures that our stacking method robustly recovers. However, the gallery also reveals more complex cases. A few systems (e.g., No.34, No.39, marked by green color) show visually net redshifted absorption, a potential signature of cool gas accreting onto the galaxy (inflow). A comprehensive analysis of these individual systems, including detailed spectral profile fitting to disentangle the complex interplay of outflows, inflows, and virial motions, is beyond the scope of this statistically-focused Letter. We will present such an analysis in a forthcoming paper.

Nevertheless, the clear presence of both dominant outflow signatures and more complex kinematic features underscores the dynamic nature of the CGM and validates our stacking approach for recovering the average, population-wide signal.

\section{The Cosmic Evolution of Outflow Velocity}
\begin{figure*}[ht!]
\includegraphics[width=1\linewidth]{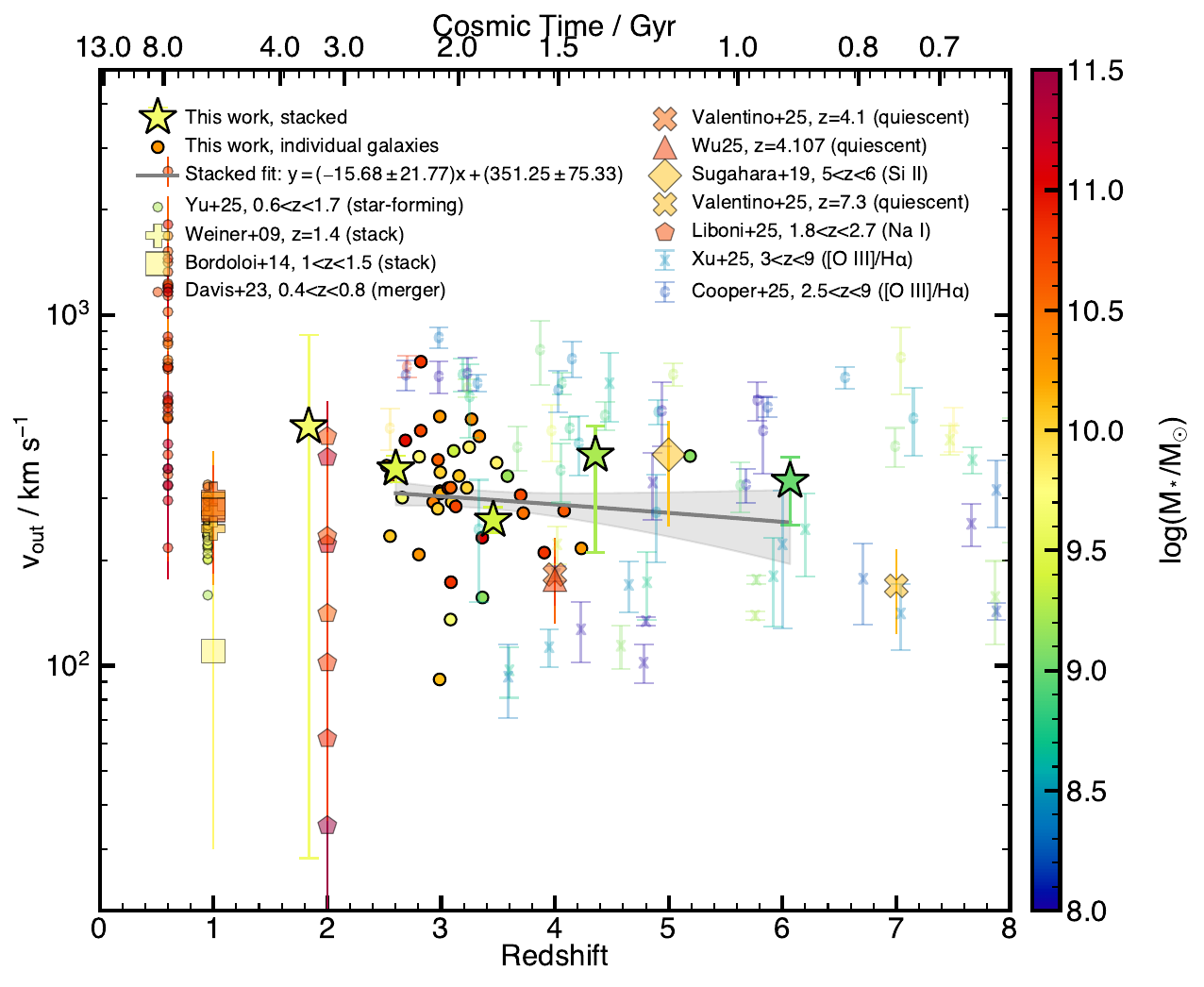}
\caption{The cosmic evolution of outflow velocity. Our measurements from the full redshift bins stacks are shown as large star symbols. The color of all data points corresponds to the weighted mean stellar mass of the respective galaxy sample, as indicated by the color bar on the right. For context, we compare our results with a compilation of literature data from \citet{Weiner2009}, \citet{Bordoloi2014},\citet{Sugahara2019}, \citet{Davis2023}, \citet{Yu2025}, \citet{Liboni2025}, \citet{Valentino2025},\citet{Cooper2025},\citet{Wu2025} and \citet{Xu2025}.
\label{fig:v_vs_z}}
\end{figure*}

Figure~\ref{fig:v_vs_z} presents another view of Figure~\ref{fig:v_vs_m}, showing the outflow velocity as a function of cosmic time. At first glance, the evolution of our measured outflow velocities (star symbols) with redshift appears complex, with no simple monotonic trend. The resulting best-fit relation is marked in the legend. However, this is primarily a consequence of selection effects, as the mean stellar mass of our stacked samples generally decreases with increasing redshift, which is clearly visualized by the color-coding of the data points. Nevertheless, this analysis highlights the unique contribution of our work: providing statistically derived outflow velocities for typical star-forming galaxies at $z > 3$ and approaching cosmic dawn.

\bibliography{sample701}{}

@ARTICLE{Wang-2019,
       author = {{Wang}, Enci and {Lilly}, Simon J. and {Pezzulli}, Gabriele and {Matthee}, Jorryt},
        title = "{On the Elevation and Suppression of Star Formation within Galaxies}",
      journal = {\apj},
         year = 2019,
        month = jun,
       volume = {877},
       number = {2},
          eid = {132},
        pages = {132},
          doi = {10.3847/1538-4357/ab1c5b},
archivePrefix = {arXiv},
       eprint = {1901.10276},
 primaryClass = {astro-ph.GA},
       adsurl = {https://ui.adsabs.harvard.edu/abs/2019ApJ...877..132W}
}

@ARTICLE{Behroozi2019,
       author = {{Behroozi}, Peter and {Wechsler}, Risa H. and {Hearin}, Andrew P. and {Conroy}, Charlie},
        title = "{UNIVERSEMACHINE: The correlation between galaxy growth and dark matter halo assembly from z = 0-10}",
      journal = {\mnras},
         year = 2019,
        month = sep,
       volume = {488},
       number = {3},
        pages = {3143-3194},
          doi = {10.1093/mnras/stz1182},
archivePrefix = {arXiv},
       eprint = {1806.07893},
 primaryClass = {astro-ph.GA},
       adsurl = {https://ui.adsabs.harvard.edu/abs/2019MNRAS.488.3143B}
}

@ARTICLE{Wu2025,
       author = {{Wu}, Po-Feng},
        title = "{Ejective Feedback as a Quenching Mechanism in the First 1.5 Billion Years of the Universe: Detection of Neutral Gas Outflow in a z = 4 Recently Quenched Galaxy}",
      journal = {\apj},
         year = 2025,
        month = jan,
       volume = {978},
       number = {2},
          eid = {131},
        pages = {131},
          doi = {10.3847/1538-4357/ad98ef},
archivePrefix = {arXiv},
       eprint = {2409.00471},
 primaryClass = {astro-ph.GA},
       adsurl = {https://ui.adsabs.harvard.edu/abs/2025ApJ...978..131W}
}

@ARTICLE{Girelli2020,
       author = {{Girelli}, G. and {Pozzetti}, L. and {Bolzonella}, M. and {Giocoli}, C. and {Marulli}, F. and {Baldi}, M.},
        title = "{The stellar-to-halo mass relation over the past 12 Gyr. I. Standard {\ensuremath{\Lambda}}CDM model}",
      journal = {\aap},
         year = 2020,
        month = feb,
       volume = {634},
          eid = {A135},
        pages = {A135},
          doi = {10.1051/0004-6361/201936329},
archivePrefix = {arXiv},
       eprint = {2001.02230},
 primaryClass = {astro-ph.CO},
       adsurl = {https://ui.adsabs.harvard.edu/abs/2020A&A...634A.135G}
}

@ARTICLE{Wang-2022a,
       author = {{Wang}, Enci and {Lilly}, Simon J.},
        title = "{Gas-phase Metallicity Profiles of Star-forming Galaxies in the Modified Accretion Disk Framework}",
      journal = {\apj},
         year = 2022,
        month = apr,
       volume = {929},
       number = {1},
          eid = {95},
        pages = {95},
          doi = {10.3847/1538-4357/ac5e31},
archivePrefix = {arXiv},
       eprint = {2201.04151},
 primaryClass = {astro-ph.GA},
       adsurl = {https://ui.adsabs.harvard.edu/abs/2022ApJ...929...95W}
}

@ARTICLE{Ma-2024,
       author = {{Ma}, Chengyu and {Wang}, Kai and {Wang}, Enci and {Peng}, Yingjie and {Jiang}, Haochen and {Yu}, Haoran and {Jia}, Cheng and {Chen}, Zeyu and {Li}, Haixin and {Kong}, Xu},
        title = "{Revisiting the Fundamental Metallicity Relation with Observation and Simulation}",
      journal = {\apjl},
         year = 2024,
        month = aug,
       volume = {971},
       number = {1},
          eid = {L14},
        pages = {L14},
          doi = {10.3847/2041-8213/ad675f},
archivePrefix = {arXiv},
       eprint = {2407.21716},
 primaryClass = {astro-ph.GA},
       adsurl = {https://ui.adsabs.harvard.edu/abs/2024ApJ...971L..14M}
}

@ARTICLE{Jia-2025,
       author = {{Jia}, Cheng and {Wang}, Enci and {Lyu}, Cheqiu and {Ma}, Chengyu and {Song}, Jie and {Chen}, Yangyao and {Wang}, Kai and {Yu}, Haoran and {Chen}, Zeyu and {Wang}, Jinyang and {Wang}, Yifan and {Kong}, Xu},
        title = "{Potential-driven Metal Cycling: JADES Census of Gas-phase Metallicity for Galaxies at 1 < z < 7}",
      journal = {\apjl},
         year = 2025,
        month = jun,
       volume = {986},
       number = {2},
          eid = {L24},
        pages = {L24},
          doi = {10.3847/2041-8213/addfd9},
archivePrefix = {arXiv},
       eprint = {2504.18820},
 primaryClass = {astro-ph.GA},
       adsurl = {https://ui.adsabs.harvard.edu/abs/2025ApJ...986L..24J}
}

@ARTICLE{Valentino2025,
       author = {{Valentino}, F. and {Heintz}, K.~E. and {Brammer}, G. and {Ito}, K. and {Kokorev}, V. and {Whitaker}, K.~E. and {Gallazzi}, A. and {de Graaff}, A. and {Weibel}, A. and {Frye}, B.~L. and {Kamieneski}, P.~S. and {Jin}, S. and {Ceverino}, D. and {Faisst}, A. and {Farcy}, M. and {Fujimoto}, S. and {Gillman}, S. and {Gottumukkala}, R. and {Hamadouche}, M. and {Harrington}, K.~C. and {Hirschmann}, M. and {Jespersen}, C.~K. and {Kakimoto}, T. and {Kubo}, M. and {Lagos}, C. d. P. and {Lee}, M. and {Magdis}, G.~E. and {Man}, A.~W.~S. and {Onodera}, M. and {Rizzo}, F. and {Shimakawa}, R. and {Setton}, D.~J. and {Tanaka}, M. and {Toft}, S. and {Wu}, P.-F. and {Zhu}, P.},
        title = "{Gas outflows in two recently quenched galaxies at z = 4 and 7}",
      journal = {\aap},
         year = 2025,
        month = jul,
       volume = {699},
          eid = {A358},
        pages = {A358},
          doi = {10.1051/0004-6361/202553908},
archivePrefix = {arXiv},
       eprint = {2503.01990},
 primaryClass = {astro-ph.GA},
       adsurl = {https://ui.adsabs.harvard.edu/abs/2025A&A...699A.358V}
}

@ARTICLE{Pollock2025,
       author = {{Pollock}, Clara L. and {Gottumukkala}, Rashmi and {Heintz}, Kasper E. and {Brammer}, Gabriel B. and {Roberts-Borsani}, Guido and {Oesch}, Pascal A. and {Witstok}, Joris and {Arellano-C{\'o}rdova}, Karla Z. and {Cullen}, Fergus and {Scholte}, Dirk and {Terp}, Chamilla and {Rowland}, Lucie and {Sneppen}, Albert and {Ito}, Kei and {Valentino}, Francesco and {Matthee}, Jorryt and {Watson}, Darach and {Toft}, Sune},
        title = "{Novel $z\sim~10$ auroral line measurements extend the gradual offset of the FMR deep into the first Gyr of cosmic time}",
      journal = {arXiv e-prints},
         year = 2025,
        month = jun,
          eid = {arXiv:2506.15779},
        pages = {arXiv:2506.15779},
          doi = {10.48550/arXiv.2506.15779},
archivePrefix = {arXiv},
       eprint = {2506.15779},
 primaryClass = {astro-ph.GA},
       adsurl = {https://ui.adsabs.harvard.edu/abs/2025arXiv250615779P}
}

@ARTICLE{Asplund2009,
       author = {{Asplund}, Martin and {Grevesse}, Nicolas and {Sauval}, A. Jacques and {Scott}, Pat},
        title = "{The Chemical Composition of the Sun}",
      journal = {\araa},
         year = 2009,
        month = sep,
       volume = {47},
       number = {1},
        pages = {481-522},
          doi = {10.1146/annurev.astro.46.060407.145222},
archivePrefix = {arXiv},
       eprint = {0909.0948},
 primaryClass = {astro-ph.SR},
       adsurl = {https://ui.adsabs.harvard.edu/abs/2009ARA&A..47..481A}
}

@ARTICLE{Kehoe2025,
       author = {{Kehoe}, Emily and {Shapley}, Alice E. and {Sanders}, Ryan L. and {Reddy}, Naveen A. and {Topping}, Michael W. and {Lam}, Natalie and {Clarke}, Leonardo and {Cullen}, Fergus and {Ellis}, Richard S. and {Forster Schreiber}, N.~M. and {Jones}, Tucker and {Khostovan}, Ali Ahmad and {McLeod}, Derek J. and {McLure}, Ross J. and {Narayanan}, Desika and {Oesch}, Pascal and {Pahl}, Anthony J.},
        title = "{The AURORA Survey: Tracing Galactic Outflows at $zrsim2.5$ with JWST/NIRSpec NUV Absorption Lines}",
      journal = {arXiv e-prints},
         year = 2025,
        month = jun,
          eid = {arXiv:2506.17381},
        pages = {arXiv:2506.17381},
          doi = {10.48550/arXiv.2506.17381},
archivePrefix = {arXiv},
       eprint = {2506.17381},
 primaryClass = {astro-ph.GA},
       adsurl = {https://ui.adsabs.harvard.edu/abs/2025arXiv250617381K}
}

@ARTICLE{Liboni2025,
       author = {{Liboni}, Caterina and {Belli}, Sirio and {Bugiani}, Letizia and {Davies}, Rebecca and {Park}, Minjung and {Conroy}, Charlie and {Emami}, Razieh and {Johnson}, Benjamin D. and {Khoram}, Amir H. and {Leja}, Joel and {Maheson}, Gabriel and {Sapori}, Matteo and {Mendel}, Trevor and {Tacchella}, Sandro and {Weinberger}, Rainer},
        title = "{Probing neutral outflows in z \raisebox{-0.5ex}\textasciitilde 2 galaxies using JWST observations of Ca II H and K absorption lines}",
      journal = {arXiv e-prints},
         year = 2025,
        month = jun,
          eid = {arXiv:2506.05470},
        pages = {arXiv:2506.05470},
          doi = {10.48550/arXiv.2506.05470},
archivePrefix = {arXiv},
       eprint = {2506.05470},
 primaryClass = {astro-ph.GA},
       adsurl = {https://ui.adsabs.harvard.edu/abs/2025arXiv250605470L}
}

@ARTICLE{deGraaff2025,
       author = {{de Graaff}, Anna and {Brammer}, Gabriel and {Weibel}, Andrea and {Lewis}, Zach and {Maseda}, Michael V. and {Oesch}, Pascal A. and {Bezanson}, Rachel and {Boogaard}, Leindert A. and {Cleri}, Nikko J. and {Cooper}, Olivia R. and {Gottumukkala}, Rashmi and {Greene}, Jenny E. and {Hirschmann}, Michaela and {Hviding}, Raphael E. and {Katz}, Harley and {Labb{\'e}}, Ivo and {Leja}, Joel and {Matthee}, Jorryt and {McConachie}, Ian and {Miller}, Tim B. and {Naidu}, Rohan P. and {Price}, Sedona H. and {Rix}, Hans-Walter and {Setton}, David J. and {Suess}, Katherine A. and {Wang}, Bingjie and {Whitaker}, Katherine E. and {Williams}, Christina C.},
        title = "{RUBIES: A complete census of the bright and red distant Universe with JWST/NIRSpec}",
      journal = {\aap},
         year = 2025,
        month = may,
       volume = {697},
          eid = {A189},
        pages = {A189},
          doi = {10.1051/0004-6361/202452186},
archivePrefix = {arXiv},
       eprint = {2409.05948},
 primaryClass = {astro-ph.GA},
       adsurl = {https://ui.adsabs.harvard.edu/abs/2025A&A...697A.189D}
}

@ARTICLE{Pessa2024,
       author = {{Pessa}, Ismael and {Wisotzki}, Lutz and {Urrutia}, Tanya and {Pharo}, John and {Augustin}, Ramona and {Bouch{\'e}}, Nicolas F. and {Feltre}, Anna and {Guo}, Yucheng and {Kozlova}, Daria and {Krajnovic}, Davor and {Kusakabe}, Haruka and {Leclercq}, Floriane and {Salas}, H{\'e}ctor and {Schaye}, Joop and {Verhamme}, Anne},
        title = "{A galactic outflow traced by its extended Mg II emission out to a {\ensuremath{\sim}}30 kpc radius in the Hubble Ultra Deep Field with MUSE}",
      journal = {\aap},
         year = 2024,
        month = nov,
       volume = {691},
          eid = {A5},
        pages = {A5},
          doi = {10.1051/0004-6361/202450547},
archivePrefix = {arXiv},
       eprint = {2408.16067},
 primaryClass = {astro-ph.GA},
       adsurl = {https://ui.adsabs.harvard.edu/abs/2024A&A...691A...5P}
}

@ARTICLE{Heintz2025,
       author = {{Heintz}, K.~E. and {Brammer}, G.~B. and {Watson}, D. and {Oesch}, P.~A. and {Keating}, L.~C. and {Hayes}, M.~J. and {Abdurro'uf} and {Arellano-C{\'o}rdova}, K.~Z. and {Carnall}, A.~C. and {Christiansen}, C.~R. and {Cullen}, F. and {Dav{\'e}}, R. and {Dayal}, P. and {Ferrara}, A. and {Finlator}, K. and {Fynbo}, J.~P.~U. and {Flury}, S.~R. and {Gelli}, V. and {Gillman}, S. and {Gottumukkala}, R. and {Gould}, K. and {Greve}, T.~R. and {Hardin}, S.~E. and {Hsiao}, T.~Y.-Y. and {Hutter}, A. and {Jakobsson}, P. and {Killi}, M. and {Khosravaninezhad}, N. and {Laursen}, P. and {Lee}, M.~M. and {Magdis}, G.~E. and {Matthee}, J. and {Naidu}, R.~P. and {Narayanan}, D. and {Pollock}, C. and {Prescott}, M.~K.~M. and {Rusakov}, V. and {Shuntov}, M. and {Sneppen}, A. and {Smit}, R. and {Tanvir}, N.~R. and {Terp}, C. and {Toft}, S. and {Valentino}, F. and {Vijayan}, A.~P. and {Weaver}, J.~R. and {Wise}, J.~H. and {Witstok}, J.},
        title = "{The JWST-PRIMAL archival survey: A JWST/NIRSpec reference sample for the physical properties and Lyman-{\ensuremath{\alpha}} absorption and emission of {\ensuremath{\sim}}600 galaxies at z = 5.0 ‑ 13.4}",
      journal = {\aap},
         year = 2025,
        month = jan,
       volume = {693},
          eid = {A60},
        pages = {A60},
          doi = {10.1051/0004-6361/202450243},
archivePrefix = {arXiv},
       eprint = {2404.02211},
 primaryClass = {astro-ph.GA},
       adsurl = {https://ui.adsabs.harvard.edu/abs/2025A&A...693A..60H}
}

@ARTICLE{Belli2024,
       author = {{Belli}, Sirio and {Park}, Minjung and {Davies}, Rebecca L. and {Mendel}, J. Trevor and {Johnson}, Benjamin D. and {Conroy}, Charlie and {Benton}, Chlo{\"e} and {Bugiani}, Letizia and {Emami}, Razieh and {Leja}, Joel and {Li}, Yijia and {Maheson}, Gabriel and {Mathews}, Elijah P. and {Naidu}, Rohan P. and {Nelson}, Erica J. and {Tacchella}, Sandro and {Terrazas}, Bryan A. and {Weinberger}, Rainer},
        title = "{Star formation shut down by multiphase gas outflow in a galaxy at a redshift of 2.45}",
      journal = {\nat},
         year = 2024,
        month = jun,
       volume = {630},
       number = {8015},
        pages = {54-58},
          doi = {10.1038/s41586-024-07412-1},
archivePrefix = {arXiv},
       eprint = {2308.05795},
 primaryClass = {astro-ph.GA},
       adsurl = {https://ui.adsabs.harvard.edu/abs/2024Natur.630...54B}
}

@ARTICLE{Davies2024,
       author = {{Davies}, Rebecca L. and {Belli}, Sirio and {Park}, Minjung and {Mendel}, J. Trevor and {Johnson}, Benjamin D. and {Conroy}, Charlie and {Benton}, Chlo{\"e} and {Bugiani}, Letizia and {Emami}, Razieh and {Leja}, Joel and {Li}, Yijia and {Maheson}, Gabriel and {Mathews}, Elijah P. and {Naidu}, Rohan P. and {Nelson}, Erica J. and {Tacchella}, Sandro and {Terrazas}, Bryan A. and {Weinberger}, Rainer},
        title = "{JWST reveals widespread AGN-driven neutral gas outflows in massive z   2 galaxies}",
      journal = {\mnras},
         year = 2024,
        month = mar,
       volume = {528},
       number = {3},
        pages = {4976-4992},
          doi = {10.1093/mnras/stae327},
archivePrefix = {arXiv},
       eprint = {2310.17939},
 primaryClass = {astro-ph.GA},
       adsurl = {https://ui.adsabs.harvard.edu/abs/2024MNRAS.528.4976D}
}

@ARTICLE{Bordoloi2014,
       author = {{Bordoloi}, R. and {Lilly}, S.~J. and {Kacprzak}, G.~G. and {Churchill}, C.~W.},
        title = "{Modeling the Distribution of Mg II Absorbers around Galaxies Using Background Galaxies and Quasars}",
      journal = {\apj},
         year = 2014,
        month = apr,
       volume = {784},
       number = {2},
          eid = {108},
        pages = {108},
          doi = {10.1088/0004-637X/784/2/108},
archivePrefix = {arXiv},
       eprint = {1211.3774},
 primaryClass = {astro-ph.CO},
       adsurl = {https://ui.adsabs.harvard.edu/abs/2014ApJ...784..108B}
}

@software{Brammer2023,
       author = {{Brammer}, Gabriel},
        title = "{grizli}",
         year = 2023,
        month = sep,
          eid = {10.5281/zenodo.8370018},
          doi = {10.5281/zenodo.8370018},
      version = {1.9.11},
    publisher = {Zenodo},
       adsurl = {https://ui.adsabs.harvard.edu/abs/2023zndo...8370018B}
}

@ARTICLE{Weaver2023,
       author = {{Weaver}, J.~R. and {Davidzon}, I. and {Toft}, S. and {Ilbert}, O. and {McCracken}, H.~J. and {Gould}, K.~M.~L. and {Jespersen}, C.~K. and {Steinhardt}, C. and {Lagos}, C.~D.~P. and {Capak}, P.~L. and {Casey}, C.~M. and {Chartab}, N. and {Faisst}, A.~L. and {Hayward}, C.~C. and {Kartaltepe}, J.~S. and {Kauffmann}, O.~B. and {Koekemoer}, A.~M. and {Kokorev}, V. and {Laigle}, C. and {Liu}, D. and {Long}, A. and {Magdis}, G.~E. and {McPartland}, C.~J.~R. and {Milvang-Jensen}, B. and {Mobasher}, B. and {Moneti}, A. and {Peng}, Y. and {Sanders}, D.~B. and {Shuntov}, M. and {Sneppen}, A. and {Valentino}, F. and {Zalesky}, L. and {Zamorani}, G.},
        title = "{COSMOS2020: The galaxy stellar mass function. The assembly and star formation cessation of galaxies at 0.2< z {\ensuremath{\leq}} 7.5}",
      journal = {\aap},
         year = 2023,
        month = sep,
       volume = {677},
          eid = {A184},
        pages = {A184},
          doi = {10.1051/0004-6361/202245581},
archivePrefix = {arXiv},
       eprint = {2212.02512},
 primaryClass = {astro-ph.GA},
       adsurl = {https://ui.adsabs.harvard.edu/abs/2023A&A...677A.184W}
}

@ARTICLE{Davis2023,
       author = {{Davis}, Julie D. and {Tremonti}, Christy A. and {Swiggum}, Cameren N. and {Moustakas}, John and {Diamond-Stanic}, Aleksandar M. and {Coil}, Alison L. and {Geach}, James E. and {Hickox}, Ryan C. and {Perrotta}, Serena and {Petter}, Grayson C. and {Rudnick}, Gregory H. and {Rupke}, David S.~N. and {Sell}, Paul H. and {Whalen}, Kelly E.},
        title = "{Extending the Dynamic Range of Galaxy Outflow Scaling Relations: Massive Compact Galaxies with Extreme Outflows}",
      journal = {\apj},
         year = 2023,
        month = jul,
       volume = {951},
       number = {2},
          eid = {105},
        pages = {105},
          doi = {10.3847/1538-4357/accbbf},
       adsurl = {https://ui.adsabs.harvard.edu/abs/2023ApJ...951..105D}
}

@ARTICLE{Cresci2023,
       author = {{Cresci}, G. and {Tozzi}, G. and {Perna}, M. and {Brusa}, M. and {Marconcini}, C. and {Marconi}, A. and {Carniani}, S. and {Brienza}, M. and {Giroletti}, M. and {Belfiore}, F. and {Ginolfi}, M. and {Mannucci}, F. and {Ulivi}, L. and {Scholtz}, J. and {Venturi}, G. and {Arribas}, S. and {{\"U}bler}, H. and {D'Eugenio}, F. and {Mingozzi}, M. and {Balmaverde}, B. and {Capetti}, A. and {Parlanti}, E. and {Zana}, T.},
        title = "{Bubbles and outflows: The novel JWST/NIRSpec view of the z = 1.59 obscured quasar XID2028}",
      journal = {\aap},
         year = 2023,
        month = apr,
       volume = {672},
          eid = {A128},
        pages = {A128},
          doi = {10.1051/0004-6361/202346001},
archivePrefix = {arXiv},
       eprint = {2301.11060},
 primaryClass = {astro-ph.GA},
       adsurl = {https://ui.adsabs.harvard.edu/abs/2023A&A...672A.128C}
}

@software{Brammer2022,
       author = {{Brammer}, Gabe},
        title = "{gbrammer/msaexp: Full working version with 2d drizzling and extraction}",
         year = 2022,
        month = nov,
          eid = {10.5281/zenodo.7299501},
          doi = {10.5281/zenodo.7299501},
      version = {0.3},
    publisher = {Zenodo},
       adsurl = {https://ui.adsabs.harvard.edu/abs/2022zndo...7299501B}
}

@ARTICLE{Baron2020,
       author = {{Baron}, Dalya and {Netzer}, Hagai and {Davies}, Ric I. and {Xavier Prochaska}, J.},
        title = "{Multiphase outflows in post-starburst E+A galaxies - II. A direct connection between the neutral and ionized outflow phases}",
      journal = {\mnras},
         year = 2020,
        month = jun,
       volume = {494},
       number = {4},
        pages = {5396-5420},
          doi = {10.1093/mnras/staa1018},
archivePrefix = {arXiv},
       eprint = {2004.04749},
 primaryClass = {astro-ph.GA},
       adsurl = {https://ui.adsabs.harvard.edu/abs/2020MNRAS.494.5396B}
}

@ARTICLE{Veilleux2020,
       author = {{Veilleux}, Sylvain and {Maiolino}, Roberto and {Bolatto}, Alberto D. and {Aalto}, Susanne},
        title = "{Cool outflows in galaxies and their implications}",
      journal = {\aapr},
         year = 2020,
        month = apr,
       volume = {28},
       number = {1},
          eid = {2},
        pages = {2},
          doi = {10.1007/s00159-019-0121-9},
archivePrefix = {arXiv},
       eprint = {2002.07765},
 primaryClass = {astro-ph.GA},
       adsurl = {https://ui.adsabs.harvard.edu/abs/2020A&ARv..28....2V}
}

@ARTICLE{ForsterSchreiber2019,
       author = {{F{\"o}rster Schreiber}, N.~M. and {{\"U}bler}, H. and {Davies}, R.~L. and {Genzel}, R. and {Wisnioski}, E. and {Belli}, S. and {Shimizu}, T. and {Lutz}, D. and {Fossati}, M. and {Herrera-Camus}, R. and {Mendel}, J.~T. and {Tacconi}, L.~J. and {Wilman}, D. and {Beifiori}, A. and {Brammer}, G.~B. and {Burkert}, A. and {Carollo}, C.~M. and {Davies}, R.~I. and {Eisenhauer}, F. and {Fabricius}, M. and {Lilly}, S.~J. and {Momcheva}, I. and {Naab}, T. and {Nelson}, E.~J. and {Price}, S.~H. and {Renzini}, A. and {Saglia}, R. and {Sternberg}, A. and {van Dokkum}, P. and {Wuyts}, S.},
        title = "{The KMOS$^{3D}$ Survey: Demographics and Properties of Galactic Outflows at z = 0.6-2.7}",
      journal = {\apj},
         year = 2019,
        month = apr,
       volume = {875},
       number = {1},
          eid = {21},
        pages = {21},
          doi = {10.3847/1538-4357/ab0ca2},
archivePrefix = {arXiv},
       eprint = {1807.04738},
 primaryClass = {astro-ph.GA},
       adsurl = {https://ui.adsabs.harvard.edu/abs/2019ApJ...875...21F}
}

@ARTICLE{Tumlinson2017,
       author = {{Tumlinson}, Jason and {Peeples}, Molly S. and {Werk}, Jessica K.},
        title = "{The Circumgalactic Medium}",
      journal = {\araa},
         year = 2017,
        month = aug,
       volume = {55},
       number = {1},
        pages = {389-432},
          doi = {10.1146/annurev-astro-091916-055240},
archivePrefix = {arXiv},
       eprint = {1709.09180},
 primaryClass = {astro-ph.GA},
       adsurl = {https://ui.adsabs.harvard.edu/abs/2017ARA&A..55..389T}
}

@ARTICLE{Carnall2017,
       author = {{Carnall}, A.~C.},
        title = "{SpectRes: A Fast Spectral Resampling Tool in Python}",
      journal = {arXiv e-prints},
         year = 2017,
        month = may,
          eid = {arXiv:1705.05165},
        pages = {arXiv:1705.05165},
          doi = {10.48550/arXiv.1705.05165},
archivePrefix = {arXiv},
       eprint = {1705.05165},
 primaryClass = {astro-ph.IM},
       adsurl = {https://ui.adsabs.harvard.edu/abs/2017arXiv170505165C}
}

@ARTICLE{Erb2015,
       author = {{Erb}, Dawn K.},
        title = "{Feedback in low-mass galaxies in the early Universe}",
      journal = {\nat},
         year = 2015,
        month = jul,
       volume = {523},
       number = {7559},
        pages = {169-176},
          doi = {10.1038/nature14454},
archivePrefix = {arXiv},
       eprint = {1507.02374},
 primaryClass = {astro-ph.GA},
       adsurl = {https://ui.adsabs.harvard.edu/abs/2015Natur.523..169E}
}

@ARTICLE{Rubin2014,
       author = {{Rubin}, Kate H.~R. and {Prochaska}, J. Xavier and {Koo}, David C. and {Phillips}, Andrew C. and {Martin}, Crystal L. and {Winstrom}, Lucas O.},
        title = "{Evidence for Ubiquitous Collimated Galactic-scale Outflows along the Star-forming Sequence at z \raisebox{-0.5ex}\textasciitilde 0.5}",
      journal = {\apj},
         year = 2014,
        month = oct,
       volume = {794},
       number = {2},
          eid = {156},
        pages = {156},
          doi = {10.1088/0004-637X/794/2/156},
archivePrefix = {arXiv},
       eprint = {1307.1476},
 primaryClass = {astro-ph.CO},
       adsurl = {https://ui.adsabs.harvard.edu/abs/2014ApJ...794..156R}
}

@ARTICLE{Martin2012,
       author = {{Martin}, Crystal L. and {Shapley}, Alice E. and {Coil}, Alison L. and {Kornei}, Katherine A. and {Bundy}, Kevin and {Weiner}, Benjamin J. and {Noeske}, Kai G. and {Schiminovich}, David},
        title = "{Demographics and Physical Properties of Gas Outflows/Inflows at 0.4 < z < 1.4}",
      journal = {\apj},
         year = 2012,
        month = dec,
       volume = {760},
       number = {2},
          eid = {127},
        pages = {127},
          doi = {10.1088/0004-637X/760/2/127},
archivePrefix = {arXiv},
       eprint = {1206.5552},
 primaryClass = {astro-ph.CO},
       adsurl = {https://ui.adsabs.harvard.edu/abs/2012ApJ...760..127M}
}

@ARTICLE{Weiner2009,
       author = {{Weiner}, Benjamin J. and {Coil}, Alison L. and {Prochaska}, Jason X. and {Newman}, Jeffrey A. and {Cooper}, Michael C. and {Bundy}, Kevin and {Conselice}, Christopher J. and {Dutton}, Aaron A. and {Faber}, S.~M. and {Koo}, David C. and {Lotz}, Jennifer M. and {Rieke}, G.~H. and {Rubin}, K.~H.~R.},
        title = "{Ubiquitous Outflows in DEEP2 Spectra of Star-Forming Galaxies at z = 1.4}",
      journal = {\apj},
         year = 2009,
        month = feb,
       volume = {692},
       number = {1},
        pages = {187-211},
          doi = {10.1088/0004-637X/692/1/187},
archivePrefix = {arXiv},
       eprint = {0804.4686},
 primaryClass = {astro-ph},
       adsurl = {https://ui.adsabs.harvard.edu/abs/2009ApJ...692..187W}
}

@ARTICLE{Somerville2008,
       author = {{Somerville}, Rachel S. and {Hopkins}, Philip F. and {Cox}, Thomas J. and {Robertson}, Brant E. and {Hernquist}, Lars},
        title = "{A semi-analytic model for the co-evolution of galaxies, black holes and active galactic nuclei}",
      journal = {\mnras},
         year = 2008,
        month = dec,
       volume = {391},
       number = {2},
        pages = {481-506},
          doi = {10.1111/j.1365-2966.2008.13805.x},
archivePrefix = {arXiv},
       eprint = {0808.1227},
 primaryClass = {astro-ph},
       adsurl = {https://ui.adsabs.harvard.edu/abs/2008MNRAS.391..481S}
}

@ARTICLE{Hopkins2008,
       author = {{Hopkins}, Philip F. and {Hernquist}, Lars and {Cox}, Thomas J. and {Kere{\v{s}}}, Du{\v{s}}an},
        title = "{A Cosmological Framework for the Co-Evolution of Quasars, Supermassive Black Holes, and Elliptical Galaxies. I. Galaxy Mergers and Quasar Activity}",
      journal = {\apjs},
         year = 2008,
        month = apr,
       volume = {175},
       number = {2},
        pages = {356-389},
          doi = {10.1086/524362},
archivePrefix = {arXiv},
       eprint = {0706.1243},
 primaryClass = {astro-ph},
       adsurl = {https://ui.adsabs.harvard.edu/abs/2008ApJS..175..356H}
}

@ARTICLE{Finlator2008,
       author = {{Finlator}, Kristian and {Dav{\'e}}, Romeel},
        title = "{The origin of the galaxy mass-metallicity relation and implications for galactic outflows}",
      journal = {\mnras},
         year = 2008,
        month = apr,
       volume = {385},
       number = {4},
        pages = {2181-2204},
          doi = {10.1111/j.1365-2966.2008.12991.x},
archivePrefix = {arXiv},
       eprint = {0704.3100},
 primaryClass = {astro-ph},
       adsurl = {https://ui.adsabs.harvard.edu/abs/2008MNRAS.385.2181F}
}

@ARTICLE{Gardner2006,
       author = {{Gardner}, Jonathan P. and {Mather}, John C. and {Clampin}, Mark and {Doyon}, Rene and {Greenhouse}, Matthew A. and {Hammel}, Heidi B. and {Hutchings}, John B. and {Jakobsen}, Peter and {Lilly}, Simon J. and {Long}, Knox S. and {Lunine}, Jonathan I. and {McCaughrean}, Mark J. and {Mountain}, Matt and {Nella}, John and {Rieke}, George H. and {Rieke}, Marcia J. and {Rix}, Hans-Walter and {Smith}, Eric P. and {Sonneborn}, George and {Stiavelli}, Massimo and {Stockman}, H.~S. and {Windhorst}, Rogier A. and {Wright}, Gillian S.},
        title = "{The James Webb Space Telescope}",
      journal = {\ssr},
         year = 2006,
        month = apr,
       volume = {123},
       number = {4},
        pages = {485-606},
          doi = {10.1007/s11214-006-8315-7},
archivePrefix = {arXiv},
       eprint = {astro-ph/0606175},
 primaryClass = {astro-ph},
       adsurl = {https://ui.adsabs.harvard.edu/abs/2006SSRv..123..485G}
}

@ARTICLE{Croton2006,
       author = {{Croton}, Darren J. and {Springel}, Volker and {White}, Simon D.~M. and {De Lucia}, G. and {Frenk}, C.~S. and {Gao}, L. and {Jenkins}, A. and {Kauffmann}, G. and {Navarro}, J.~F. and {Yoshida}, N.},
        title = "{The many lives of active galactic nuclei: cooling flows, black holes and the luminosities and colours of galaxies}",
      journal = {\mnras},
         year = 2006,
        month = jan,
       volume = {365},
       number = {1},
        pages = {11-28},
          doi = {10.1111/j.1365-2966.2005.09675.x},
archivePrefix = {arXiv},
       eprint = {astro-ph/0508046},
 primaryClass = {astro-ph},
       adsurl = {https://ui.adsabs.harvard.edu/abs/2006MNRAS.365...11C}
}

@ARTICLE{Veilleux2005,
       author = {{Veilleux}, Sylvain and {Cecil}, Gerald and {Bland-Hawthorn}, Joss},
        title = "{Galactic Winds}",
      journal = {\araa},
         year = 2005,
        month = sep,
       volume = {43},
       number = {1},
        pages = {769-826},
          doi = {10.1146/annurev.astro.43.072103.150610},
archivePrefix = {arXiv},
       eprint = {astro-ph/0504435},
 primaryClass = {astro-ph},
       adsurl = {https://ui.adsabs.harvard.edu/abs/2005ARA&A..43..769V}
}

@ARTICLE{Rupke2005,
       author = {{Rupke}, David S. and {Veilleux}, Sylvain and {Sanders}, D.~B.},
        title = "{Outflows in Infrared-Luminous Starbursts at z < 0.5. II. Analysis and Discussion}",
      journal = {\apjs},
         year = 2005,
        month = sep,
       volume = {160},
       number = {1},
        pages = {115-148},
          doi = {10.1086/432889},
archivePrefix = {arXiv},
       eprint = {astro-ph/0506611},
 primaryClass = {astro-ph},
       adsurl = {https://ui.adsabs.harvard.edu/abs/2005ApJS..160..115R}
}

@ARTICLE{DiMatteo2005,
       author = {{Di Matteo}, Tiziana and {Springel}, Volker and {Hernquist}, Lars},
        title = "{Energy input from quasars regulates the growth and activity of black holes and their host galaxies}",
      journal = {\nat},
         year = 2005,
        month = feb,
       volume = {433},
       number = {7026},
        pages = {604-607},
          doi = {10.1038/nature03335},
archivePrefix = {arXiv},
       eprint = {astro-ph/0502199},
 primaryClass = {astro-ph},
       adsurl = {https://ui.adsabs.harvard.edu/abs/2005Natur.433..604D}
}

@ARTICLE{Tremonti2004,
       author = {{Tremonti}, Christy A. and {Heckman}, Timothy M. and {Kauffmann}, Guinevere and {Brinchmann}, Jarle and {Charlot}, St{\'e}phane and {White}, Simon D.~M. and {Seibert}, Mark and {Peng}, Eric W. and {Schlegel}, David J. and {Uomoto}, Alan and {Fukugita}, Masataka and {Brinkmann}, Jon},
        title = "{The Origin of the Mass-Metallicity Relation: Insights from 53,000 Star-forming Galaxies in the Sloan Digital Sky Survey}",
      journal = {\apj},
         year = 2004,
        month = oct,
       volume = {613},
       number = {2},
        pages = {898-913},
          doi = {10.1086/423264},
archivePrefix = {arXiv},
       eprint = {astro-ph/0405537},
 primaryClass = {astro-ph},
       adsurl = {https://ui.adsabs.harvard.edu/abs/2004ApJ...613..898T}
}

@ARTICLE{Chabrier2003,
       author = {{Chabrier}, Gilles},
        title = "{Galactic Stellar and Substellar Initial Mass Function}",
      journal = {\pasp},
         year = 2003,
        month = jul,
       volume = {115},
       number = {809},
        pages = {763-795},
          doi = {10.1086/376392},
archivePrefix = {arXiv},
       eprint = {astro-ph/0304382},
 primaryClass = {astro-ph},
       adsurl = {https://ui.adsabs.harvard.edu/abs/2003PASP..115..763C}
}

@ARTICLE{Pettini2001,
       author = {{Pettini}, Max and {Shapley}, Alice E. and {Steidel}, Charles C. and {Cuby}, Jean-Gabriel and {Dickinson}, Mark and {Moorwood}, Alan F.~M. and {Adelberger}, Kurt L. and {Giavalisco}, Mauro},
        title = "{The Rest-Frame Optical Spectra of Lyman Break Galaxies: Star Formation, Extinction, Abundances, and Kinematics}",
      journal = {\apj},
         year = 2001,
        month = jun,
       volume = {554},
       number = {2},
        pages = {981-1000},
          doi = {10.1086/321403},
archivePrefix = {arXiv},
       eprint = {astro-ph/0102456},
 primaryClass = {astro-ph},
       adsurl = {https://ui.adsabs.harvard.edu/abs/2001ApJ...554..981P}
}

@ARTICLE{Churchill2000,
       author = {{Churchill}, Christopher W. and {Mellon}, Richard R. and {Charlton}, Jane C. and {Jannuzi}, Buell T. and {Kirhakos}, Sofia and {Steidel}, Charles C. and {Schneider}, Donald P.},
        title = "{Low- and High-Ionization Absorption Properties of Mg II Absorption-selected Galaxies at Intermediate Redshifts. II. Taxonomy, Kinematics, and Galaxies}",
      journal = {\apj},
         year = 2000,
        month = nov,
       volume = {543},
       number = {2},
        pages = {577-598},
          doi = {10.1086/317120},
archivePrefix = {arXiv},
       eprint = {astro-ph/0005586},
 primaryClass = {astro-ph},
       adsurl = {https://ui.adsabs.harvard.edu/abs/2000ApJ...543..577C}
}

@ARTICLE{Heckman2000,
       author = {{Heckman}, Timothy M. and {Lehnert}, Matthew D. and {Strickland}, David K. and {Armus}, Lee},
        title = "{Absorption-Line Probes of Gas and Dust in Galactic Superwinds}",
      journal = {\apjs},
         year = 2000,
        month = aug,
       volume = {129},
       number = {2},
        pages = {493-516},
          doi = {10.1086/313421},
archivePrefix = {arXiv},
       eprint = {astro-ph/0002526},
 primaryClass = {astro-ph},
       adsurl = {https://ui.adsabs.harvard.edu/abs/2000ApJS..129..493H}
}

@ARTICLE{Silk1998,
       author = {{Silk}, Joseph and {Rees}, Martin J.},
        title = "{Quasars and galaxy formation}",
      journal = {\aap},
         year = 1998,
        month = mar,
       volume = {331},
        pages = {L1-L4},
          doi = {10.48550/arXiv.astro-ph/9801013},
archivePrefix = {arXiv},
       eprint = {astro-ph/9801013},
 primaryClass = {astro-ph},
       adsurl = {https://ui.adsabs.harvard.edu/abs/1998A&A...331L...1S}
}

@ARTICLE{Savage1996,
       author = {{Savage}, Blair D. and {Sembach}, Kenneth R.},
        title = "{Interstellar Abundances from Absorption-Line Observations with the Hubble Space Telescope}",
      journal = {\araa},
         year = 1996,
        month = jan,
       volume = {34},
        pages = {279-330},
          doi = {10.1146/annurev.astro.34.1.279},
       adsurl = {https://ui.adsabs.harvard.edu/abs/1996ARA&A..34..279S}
}

@ARTICLE{Bergeron1986,
       author = {{Bergeron}, J. and {Stasi{\'n}ska}, G.},
        title = "{Absorption line systems in QSO spectra : properties derived from observations and from photoionization models.}",
      journal = {\aap},
         year = 1986,
        month = nov,
       volume = {169},
        pages = {1-3},
       adsurl = {https://ui.adsabs.harvard.edu/abs/1986A&A...169....1B}
}

@ARTICLE{Rubin2010,
       author = {{Rubin}, Kate H.~R. and {Weiner}, Benjamin J. and {Koo}, David C. and {Martin}, Crystal L. and {Prochaska}, J. Xavier and {Coil}, Alison L. and {Newman}, Jeffrey A.},
        title = "{The Persistence of Cool Galactic Winds in High Stellar Mass Galaxies between z \raisebox{-0.5ex}\textasciitilde 1.4 and \raisebox{-0.5ex}\textasciitilde1}",
      journal = {\apj},
         year = 2010,
        month = aug,
       volume = {719},
       number = {2},
        pages = {1503-1525},
          doi = {10.1088/0004-637X/719/2/1503},
archivePrefix = {arXiv},
       eprint = {0912.2343},
 primaryClass = {astro-ph.CO},
       adsurl = {https://ui.adsabs.harvard.edu/abs/2010ApJ...719.1503R}
}

@ARTICLE{Chen2010,
       author = {{Chen}, Yan-Mei and {Tremonti}, Christy A. and {Heckman}, Timothy M. and {Kauffmann}, Guinevere and {Weiner}, Benjamin J. and {Brinchmann}, Jarle and {Wang}, Jing},
        title = "{Absorption-line Probes of the Prevalence and Properties of Outflows in Present-day Star-forming Galaxies}",
      journal = {\aj},
         year = 2010,
        month = aug,
       volume = {140},
       number = {2},
        pages = {445-461},
          doi = {10.1088/0004-6256/140/2/445},
archivePrefix = {arXiv},
       eprint = {1003.5425},
 primaryClass = {astro-ph.GA},
       adsurl = {https://ui.adsabs.harvard.edu/abs/2010AJ....140..445C}
}

@article{Yu2025,
author = {Haoran Yu and Enci Wang and Zeyu Chen and Céline Péroux and Hu Zou and Zhicheng He and Huiyuan Wang and Cheqiu Lyu and Cheng Jia and Chengyu Ma and Xu Kong},
title = {Stellar feedback drives the baryon deficiency in low-mass galaxies},
year = {2025},
url = {https://arxiv.org/abs/2512.05584},
archivePrefix = {arXiv},
eprint = {2512.05584},
}

@ARTICLE{Clarke2025,
       author = {{Clarke}, Leonardo and {Shapley}, Alice E. and {Lam}, Natalie and {Topping}, Michael W. and {Brammer}, Gabriel B. and {Sanders}, Ryan L. and {Reddy}, Naveen A. and {Karthikeyan}, Shreya},
        title = "{The Star-forming Main Sequence and Bursty Star-formation Histories at $z>1.4$ in JADES and AURORA}",
      journal = {arXiv e-prints},
         year = 2025,
        month = oct,
          eid = {arXiv:2510.06681},
        pages = {arXiv:2510.06681},
          doi = {10.48550/arXiv.2510.06681},
archivePrefix = {arXiv},
       eprint = {2510.06681},
 primaryClass = {astro-ph.GA},
       adsurl = {https://ui.adsabs.harvard.edu/abs/2025arXiv251006681C}
}

@ARTICLE{Lyu2025,
       author = {{Lyu}, Cheqiu and {Wang}, Enci and {Wang}, Junxian and {Jia}, Cheng and {Song}, Jie and {Chen}, Yangyao and {Chen}, Zeyu and {Yu}, Haoran and {Ma}, Chengyu and {Wang}, Jinyang and {Wang}, Yifan and {Kong}, Xu},
        title = "{Central Concentration and Escape of Ionizing Photons in Galaxies at the Epoch of Reionization}",
      journal = {\apjl},
         year = 2025,
        month = aug,
       volume = {988},
       number = {2},
          eid = {L72},
        pages = {L72},
          doi = {10.3847/2041-8213/adf0f9},
archivePrefix = {arXiv},
       eprint = {2507.16131},
 primaryClass = {astro-ph.GA},
       adsurl = {https://ui.adsabs.harvard.edu/abs/2025ApJ...988L..72L}
}

@ARTICLE{Glazer2025,
       author = {{Glazer}, Kelsey S. and {Jones}, Tucker and {Chen}, Yuguang and {Sanders}, Ryan L. and {Brada{\v{c}}}, Maru{\v{s}}a and {Pahl}, Anthony J. and {Shapley}, Alice E. and {Ellis}, Richard S. and {Topping}, Michael W. and {Reddy}, Naveen A.},
        title = "{Stacking PANCAKEZ: Spectroscopic Analysis with NIRSpec Stacks in the Epoch of Reionization. Weak Interstellar Medium Absorption and Implications for Ionizing Photon Escape at z {\ensuremath{\sim}} 7}",
      journal = {\apj},
         year = 2025,
        month = oct,
       volume = {992},
       number = {2},
          eid = {191},
        pages = {191},
          doi = {10.3847/1538-4357/ae0194},
archivePrefix = {arXiv},
       eprint = {2504.21080},
 primaryClass = {astro-ph.GA},
       adsurl = {https://ui.adsabs.harvard.edu/abs/2025ApJ...992..191G}
}

@ARTICLE{Looser2025,
       author = {{Looser}, Tobias J. and {D'Eugenio}, Francesco and {Maiolino}, Roberto and {Tacchella}, Sandro and {Curti}, Mirko and {Arribas}, Santiago and {Baker}, William M. and {Baum}, Stefi and {Bonaventura}, Nina and {Boyett}, Kristan and {Bunker}, Andrew J. and {Carniani}, Stefano and {Charlot}, Stephane and {Chevallard}, Jacopo and {Curtis-Lake}, Emma and {Lola Danhaive}, A. and {Eisenstein}, Daniel J. and {de Graaff}, Anna and {Hainline}, Kevin and {Ji}, Zhiyuan and {Johnson}, Benjamin D. and {Kumari}, Nimisha and {Nelson}, Erica and {Parlanti}, Eleonora and {Rix}, Hans-Walter and {Robertson}, Brant and {Del Pino}, Bruno Rodr{\'\i}guez and {Sandles}, Lester and {Scholtz}, Jan and {Smit}, Renske and {Stark}, Daniel P. and {{\"U}bler}, Hannah and {Williams}, Christina C. and {Willott}, Chris and {Witstok}, Joris},
        title = "{JADES: Differing assembly histories of galaxies: Observational evidence for bursty star formation histories and (mini-)quenching in the first billion years of the Universe}",
      journal = {\aap},
         year = 2025,
        month = may,
       volume = {697},
          eid = {A88},
        pages = {A88},
          doi = {10.1051/0004-6361/202347102},
archivePrefix = {arXiv},
       eprint = {2306.02470},
 primaryClass = {astro-ph.GA},
       adsurl = {https://ui.adsabs.harvard.edu/abs/2025A&A...697A..88L}
}

@ARTICLE{Wang2021,
       author = {{Wang}, Enci and {Lilly}, Simon J.},
        title = "{Gas-phase Metallicity as a Diagnostic of the Drivers of Star Formation on Different Spatial Scales}",
      journal = {\apj},
         year = 2021,
        month = apr,
       volume = {910},
       number = {2},
          eid = {137},
        pages = {137},
          doi = {10.3847/1538-4357/abe413},
archivePrefix = {arXiv},
       eprint = {2009.01935},
 primaryClass = {astro-ph.GA},
       adsurl = {https://ui.adsabs.harvard.edu/abs/2021ApJ...910..137W}
}

@ARTICLE{Lyu2025a,
       author = {{Lyu}, Cheqiu and {Wang}, Enci and {Zhang}, Hongxin and {Peng}, Yingjie and {Wang}, Xin and {Li}, Haixin and {Ma}, Chengyu and {Yu}, Haoran and {Chen}, Zeyu and {Jia}, Cheng and {Kong}, Xu},
        title = "{Dominant Role of Coplanar Inflows in Driving Disk Evolution Revealed by Gas-phase Metallicity Gradients}",
      journal = {\apjl},
         year = 2025,
        month = mar,
       volume = {981},
       number = {1},
          eid = {L6},
        pages = {L6},
          doi = {10.3847/2041-8213/adb4ed},
archivePrefix = {arXiv},
       eprint = {2502.12409},
 primaryClass = {astro-ph.GA},
       adsurl = {https://ui.adsabs.harvard.edu/abs/2025ApJ...981L...6L}
}

@ARTICLE{Chen2025a,
       author = {{Chen}, Zeyu and {Wang}, Enci and {Zou}, Hu and {Zou}, Siwei and {Gao}, Yang and {Wang}, Huiyuan and {Yu}, Haoran and {Jia}, Cheng and {Li}, Haixin and {Ma}, Chengyu and {Yao}, Yao and {Ding}, Weiyu and {Zhu}, Runyu},
        title = "{The Circumgalactic Medium Traced by Mg II Absorption with DESI: Dependence on Galaxy Stellar Mass, Star Formation Rate, and Azimuthal Angle}",
      journal = {\apj},
         year = 2025,
        month = mar,
       volume = {981},
       number = {1},
          eid = {81},
        pages = {81},
          doi = {10.3847/1538-4357/ada942},
archivePrefix = {arXiv},
       eprint = {2411.08485},
 primaryClass = {astro-ph.GA},
       adsurl = {https://ui.adsabs.harvard.edu/abs/2025ApJ...981...81C}
}

@ARTICLE{Chen2025b,
       author = {{Chen}, Zeyu and {Wang}, Enci and {Zou}, Hu and {Yu}, Haoran and {He}, Zhicheng and {Wang}, Huiyuan and {Gao}, Yang and {Lyu}, Cheqiu and {Jia}, Cheng and {Ma}, Chengyu and {Ding}, Weiyu and {Zhu}, Runyu and {Kong}, Xu},
        title = "{The Cosmic Evolution and Spatial Distribution of Multiphase Gas Associated with QSOs}",
      journal = {\apjl},
         year = 2025,
        month = jul,
       volume = {988},
       number = {1},
          eid = {L39},
        pages = {L39},
          doi = {10.3847/2041-8213/ade545},
archivePrefix = {arXiv},
       eprint = {2505.11919},
 primaryClass = {astro-ph.GA},
       adsurl = {https://ui.adsabs.harvard.edu/abs/2025ApJ...988L..39C}
}

@ARTICLE{Dekel2023,
       author = {{Dekel}, Avishai and {Sarkar}, Kartick C. and {Birnboim}, Yuval and {Mandelker}, Nir and {Li}, Zhaozhou},
        title = "{Efficient formation of massive galaxies at cosmic dawn by feedback-free starbursts}",
      journal = {\mnras},
         year = 2023,
        month = aug,
       volume = {523},
       number = {3},
        pages = {3201-3218},
          doi = {10.1093/mnras/stad1557},
archivePrefix = {arXiv},
       eprint = {2303.04827},
 primaryClass = {astro-ph.GA},
       adsurl = {https://ui.adsabs.harvard.edu/abs/2023MNRAS.523.3201D}
}

@ARTICLE{Li2024,
       author = {{Li}, Zhaozhou and {Dekel}, Avishai and {Sarkar}, Kartick C. and {Aung}, Han and {Giavalisco}, Mauro and {Mandelker}, Nir and {Tacchella}, Sandro},
        title = "{Feedback-free starbursts at cosmic dawn: Observable predictions for JWST}",
      journal = {\aap},
         year = 2024,
        month = oct,
       volume = {690},
          eid = {A108},
        pages = {A108},
          doi = {10.1051/0004-6361/202348727},
archivePrefix = {arXiv},
       eprint = {2311.14662},
 primaryClass = {astro-ph.GA},
       adsurl = {https://ui.adsabs.harvard.edu/abs/2024A&A...690A.108L}
}

@ARTICLE{Puspitarini2012,
       author = {{Puspitarini}, L. and {Lallement}, R.},
        title = "{Distance to northern high-latitude HI shells}",
      journal = {\aap},
         year = 2012,
        month = sep,
       volume = {545},
          eid = {A21},
        pages = {A21},
          doi = {10.1051/0004-6361/201219284},
archivePrefix = {arXiv},
       eprint = {1207.5353},
 primaryClass = {astro-ph.GA},
       adsurl = {https://ui.adsabs.harvard.edu/abs/2012A&A...545A..21P}
}

@ARTICLE{Sugahara2019,
       author = {{Sugahara}, Yuma and {Ouchi}, Masami and {Harikane}, Yuichi and {Bouch{\'e}}, Nicolas and {Mitchell}, Peter D. and {Blaizot}, J{\'e}r{\'e}my},
        title = "{Fast Outflows Identified in Early Star-forming Galaxies at z = 5-6}",
      journal = {\apj},
         year = 2019,
        month = nov,
       volume = {886},
       number = {1},
          eid = {29},
        pages = {29},
          doi = {10.3847/1538-4357/ab49fe},
archivePrefix = {arXiv},
       eprint = {1904.03106},
 primaryClass = {astro-ph.GA},
       adsurl = {https://ui.adsabs.harvard.edu/abs/2019ApJ...886...29S}
}

@ARTICLE{Carniani2024,
       author = {{Carniani}, Stefano and {Venturi}, Giacomo and {Parlanti}, Eleonora and {de Graaff}, Anna and {Maiolino}, Roberto and {Arribas}, Santiago and {Bonaventura}, Nina and {Boyett}, Kristan and {Bunker}, Andrew J. and {Cameron}, Alex J. and {Charlot}, Stephane and {Chevallard}, Jacopo and {Curti}, Mirko and {Curtis-Lake}, Emma and {Eisenstein}, Daniel J. and {Giardino}, Giovanna and {Hausen}, Ryan and {Kumari}, Nimisha and {Maseda}, Michael V. and {Nelson}, Erica and {Perna}, Michele and {Rix}, Hans-Walter and {Robertson}, Brant and {Del Pino}, Bruno Rodr{\'\i}guez and {Sandles}, Lester and {Scholtz}, Jan and {Simmonds}, Charlotte and {Smit}, Renske and {Tacchella}, Sandro and {{\"U}bler}, Hannah and {Williams}, Christina C. and {Willott}, Chris and {Witstok}, Joris},
        title = "{JADES: The incidence rate and properties of galactic outflows in low-mass galaxies across 3 < z < 9}",
      journal = {\aap},
         year = 2024,
        month = may,
       volume = {685},
          eid = {A99},
        pages = {A99},
          doi = {10.1051/0004-6361/202347230},
archivePrefix = {arXiv},
       eprint = {2306.11801},
 primaryClass = {astro-ph.GA},
       adsurl = {https://ui.adsabs.harvard.edu/abs/2024A&A...685A..99C}
}

@ARTICLE{Saldana-Lopez2025,
       author = {{Saldana-Lopez}, A. and {Chisholm}, J. and {Gazagnes}, S. and {Endsley}, R. and {Hayes}, M.~J. and {Berg}, D.~A. and {Finkelstein}, S.~L. and {Flury}, S.~R. and {Guseva}, N.~G. and {Henry}, A. and {Izotov}, Y.~I. and {Lambrides}, E. and {Marques-Chaves}, R. and {Richardson}, C.~T.},
        title = "{Feedback and dynamical masses in high-z galaxies: the advent of high-resolution NIRSpec spectroscopy}",
      journal = {\mnras},
         year = 2025,
        month = nov,
       volume = {544},
       number = {1},
        pages = {132-151},
          doi = {10.1093/mnras/staf1680},
archivePrefix = {arXiv},
       eprint = {2501.17145},
 primaryClass = {astro-ph.GA},
       adsurl = {https://ui.adsabs.harvard.edu/abs/2025MNRAS.544..132S}
}

@ARTICLE{Cooper2025,
       author = {{Cooper}, Ryan A. and {Caputi}, Karina I. and {Iani}, Edoardo and {Rinaldi}, Pierluigi and {Desprez}, Guillaume and {Navarro-Carrera}, Rafael},
        title = "{High-velocity Outflows in [O III] Emitters at 2.5 < z < 9 from JWST NIRSpec Medium-resolution Spectroscopy}",
      journal = {\apj},
         year = 2025,
        month = nov,
       volume = {994},
       number = {1},
          eid = {102},
        pages = {102},
          doi = {10.3847/1538-4357/ae0580},
archivePrefix = {arXiv},
       eprint = {2502.18310},
 primaryClass = {astro-ph.GA},
       adsurl = {https://ui.adsabs.harvard.edu/abs/2025ApJ...994..102C}
}

@ARTICLE{Birkin2025,
       author = {{Birkin}, Jack E. and {Spilker}, Justin S. and {Herrera-Camus}, Rodrigo and {Davies}, Rebecca L. and {Lee}, Lilian L. and {Aravena}, Manuel and {Assef}, Roberto J. and {Barcos-Mu{\~n}oz}, Loreto and {Bolatto}, Alberto and {Diaz-Santos}, Tanio and {Faisst}, Andreas L. and {Ferrara}, Andrea and {Fisher}, Deanne B. and {Gonz{\'a}lez-L{\'o}pez}, Jorge and {Ikeda}, Ryota and {Knudsen}, Kirsten and {Li}, Juno and {Li}, Yuan and {de Looze}, Ilse and {Lutz}, Dieter and {Mitsuhashi}, Ikki and {Posses}, Ana and {Rela{\~n}o}, Monica and {Solimano}, Manuel and {Tadaki}, Ken-ichi and {Villanueva}, Vicente},
        title = "{The ALMA-CRISTAL Survey: Weak Evidence for Star-formation-driven Outflows in z {\ensuremath{\sim}} 5 Main-sequence Galaxies}",
      journal = {\apj},
         year = 2025,
        month = jun,
       volume = {985},
       number = {2},
          eid = {243},
        pages = {243},
          doi = {10.3847/1538-4357/adced3},
archivePrefix = {arXiv},
       eprint = {2504.17877},
 primaryClass = {astro-ph.GA},
       adsurl = {https://ui.adsabs.harvard.edu/abs/2025ApJ...985..243B}
}

@ARTICLE{Xu2025,
       author = {{Xu}, Yi and {Ouchi}, Masami and {Nakajima}, Kimihiko and {Harikane}, Yuichi and {Isobe}, Yuki and {Ono}, Yoshiaki and {Umeda}, Hiroya and {Zhang}, Yechi},
        title = "{Stellar- and AGN-driven Outflows in JWST Galaxies at z = 3{\textendash}9: More Frequent, Wider Opening Angles, and Mostly Bounded}",
      journal = {\apj},
         year = 2025,
        month = may,
       volume = {984},
       number = {2},
          eid = {182},
        pages = {182},
          doi = {10.3847/1538-4357/adc733},
archivePrefix = {arXiv},
       eprint = {2310.06614},
 primaryClass = {astro-ph.GA},
       adsurl = {https://ui.adsabs.harvard.edu/abs/2025ApJ...984..182X}
}

@ARTICLE{Morton2003,
       author = {{Morton}, Donald C.},
        title = "{Atomic Data for Resonance Absorption Lines. III. Wavelengths Longward of the Lyman Limit for the Elements Hydrogen to Gallium}",
      journal = {\apjs},
         year = 2003,
        month = nov,
       volume = {149},
       number = {1},
        pages = {205-238},
          doi = {10.1086/377639},
       adsurl = {https://ui.adsabs.harvard.edu/abs/2003ApJS..149..205M}
}

@ARTICLE{Lan17,
       author = {{Lan}, Ting-Wen and {Fukugita}, Masataka},
        title = "{Mg II Absorbers: Metallicity Evolution and Cloud Morphology}",
      journal = {\apj},
         year = 2017,
        month = dec,
       volume = {850},
       number = {2},
          eid = {156},
        pages = {156},
          doi = {10.3847/1538-4357/aa93eb},
archivePrefix = {arXiv},
       eprint = {1707.09830},
 primaryClass = {astro-ph.GA},
       adsurl = {https://ui.adsabs.harvard.edu/abs/2017ApJ...850..156L}
}

@ARTICLE{Weldon2024,
       author = {{Weldon}, Andrew and {Reddy}, Naveen A. and {Coil}, Alison L. and {Shapley}, Alice E. and {Siana}, Brian and {Price}, Sedona H. and {Kriek}, Mariska and {Mobasher}, Bahram and {Song}, Zhiyuan and {Wozniak}, Michael A.},
        title = "{The MOSDEF survey: properties of warm ionized outflows at z = 1.4-3.8}",
      journal = {\mnras},
         year = 2024,
        month = jul,
       volume = {531},
       number = {4},
        pages = {4560-4576},
          doi = {10.1093/mnras/stae1428},
archivePrefix = {arXiv},
       eprint = {2404.05725},
 primaryClass = {astro-ph.GA},
       adsurl = {https://ui.adsabs.harvard.edu/abs/2024MNRAS.531.4560W}
}

@ARTICLE{Virtanen2020,
       author = {{Virtanen}, Pauli and {Gommers}, Ralf and {Oliphant}, Travis E. and {Haberland}, Matt and {Reddy}, Tyler and {Cournapeau}, David and {Burovski}, Evgeni and {Peterson}, Pearu and {Weckesser}, Warren and {Bright}, Jonathan and {van der Walt}, St{\'e}fan J. and {Brett}, Matthew and {Wilson}, Joshua and {Millman}, K. Jarrod and {Mayorov}, Nikolay and {Nelson}, Andrew R.~J. and {Jones}, Eric and {Kern}, Robert and {Larson}, Eric and {Carey}, C.~J. and {Polat}, {\.I}lhan and {Feng}, Yu and {Moore}, Eric W. and {VanderPlas}, Jake and {Laxalde}, Denis and {Perktold}, Josef and {Cimrman}, Robert and {Henriksen}, Ian and {Quintero}, E.~A. and {Harris}, Charles R. and {Archibald}, Anne M. and {Ribeiro}, Ant{\^o}nio H. and {Pedregosa}, Fabian and {van Mulbregt}, Paul and {SciPy 1.  0 Contributors}},
        title = "{SciPy 1.0: fundamental algorithms for scientific computing in Python}",
      journal = {Nature Medicine},
         year = 2020,
        month = feb,
       volume = {17},
        pages = {261-272},
          doi = {10.1038/s41592-019-0686-2},
archivePrefix = {arXiv},
       eprint = {1907.10121},
 primaryClass = {cs.MS},
       adsurl = {https://ui.adsabs.harvard.edu/abs/2020NatMe..17..261V}
}

@ARTICLE{Bryan1998,
       author = {{Bryan}, Greg L. and {Norman}, Michael L.},
        title = "{Statistical Properties of X-Ray Clusters: Analytic and Numerical Comparisons}",
      journal = {\apj},
         year = 1998,
        month = mar,
       volume = {495},
       number = {1},
        pages = {80-99},
          doi = {10.1086/305262},
archivePrefix = {arXiv},
       eprint = {astro-ph/9710107},
 primaryClass = {astro-ph},
       adsurl = {https://ui.adsabs.harvard.edu/abs/1998ApJ...495...80B}
}

@ARTICLE{Moster2013,
       author = {{Moster}, Benjamin P. and {Naab}, Thorsten and {White}, Simon D.~M.},
        title = "{Galactic star formation and accretion histories from matching galaxies to dark matter haloes}",
      journal = {\mnras},
         year = 2013,
        month = feb,
       volume = {428},
       number = {4},
        pages = {3121-3138},
          doi = {10.1093/mnras/sts261},
archivePrefix = {arXiv},
       eprint = {1205.5807},
 primaryClass = {astro-ph.CO},
       adsurl = {https://ui.adsabs.harvard.edu/abs/2013MNRAS.428.3121M}
}
\bibliographystyle{aasjournalv7}
\end{document}